\documentclass[manuscript,screen,natbib=false]{acmart}

\usepackage[
  backend=biber,
  style=acmnumeric
]{biblatex}

\DeclareUnicodeCharacter{2002}{ }
\DeclareUnicodeCharacter{2003}{ }
\DeclareUnicodeCharacter{202A}{}
\DeclareUnicodeCharacter{202C}{}
\AtBeginDocument{%
  }

\setcopyright{none}
\copyrightyear{2025}
\acmYear{2025}
\acmDOI{}

\usepackage{pifont}   
\usepackage{wasysym}  
\usepackage{booktabs} 
\usepackage{tabularx} 
\usepackage{graphicx}
\DeclareUnicodeCharacter{03BA}{\ensuremath{\kappa}}
\DeclareUnicodeCharacter{03A6}{\ensuremath{\Phi}}
\usepackage{caption}
\usepackage{threeparttable}
\usepackage{tikz}
\usetikzlibrary{positioning,arrows.meta}

\newcommand{\cmark}{\ding{51}}      
\newcommand{\xmark}{\ding{55}}      
\newcommand{\pmark}{\LEFTcircle}    

\begin{document}

\title{On Understanding, Identifying, and
Mitigating Vulnerabilities in Agentic Large Language Models}
\titlenote{This is the authors' preprint version. This manuscript is currently under review at ACM Computing Surveys (CSUR).}

\author{Md Jafrin Hossain}
\email{mhoss098@fiu.edu}
\affiliation{%
  \institution{Florida International University}
  \city{Miami}
  \state{Florida}
  \country{USA}
}

\author{Mohammad Arif Hossain}
\email{Mohammad.Hossain@mtsu.edu}
\affiliation{%
  \institution{Middle Tennessee State University}
  \city{Murfreesboro}
  \state{Tennessee}
  \country{USA}
}

\author{Nirwan Ansari}
\email{nirwan.ansari@njit.edu}
\affiliation{%
  \institution{New Jersey Institute of Technology}
  \city{Newark}
  \state{NJ}
  \country{USA}
}

\begin{abstract}
Large Language Models (LLMs) have undergone a shift from stateless 
conversational interfaces to autonomous agents capable of multi-step 
planning, tool invocation, code execution, and maintaining persistent memory. When these agents operate with real-world privileges---calling APIs, modifying files, and querying databases---a compromised reasoning step can trigger unauthorized data access, irreversible state changes, or cascading failures, yet the security research community has not kept pace. To quantify the state of the field, we conducted a systematic literature review under PRISMA 2020 guidelines across six databases, screening 743 records and retaining 85 papers (2023--2025) on agentic LLM security. Attack research outpaces defense work by 3.9:1. Perception-layer vulnerabilities (prompt injection, jailbreaking, adversarial perturbations) 
dominate, accounting for 66\% of papers, while action-layer vulnerabilities (tool misuse, code injection, sandbox escape) appear in only 4.7\%, misaligned with real-world risk. Code execution security accounts for 3.5\%, and tool-augmented agents 12\%. We contribute a four-layer taxonomy mapping 13 vulnerability types across perception, brain, action, and interaction layers, and identify seven open problems centered on containment. Agentic LLM insecurity stems from architectural coupling, where weak isolation allows vulnerabilities to propagate across layers.
\end{abstract}

\ccsdesc[500]{Security and privacy~Human and societal aspects of security and privacy}
\ccsdesc[500]{Computing methodologies~Artificial intelligence}

\keywords{Agentic AI, Large Language Models, LLM Agents, Vulnerability, Security, Prompt Injection, Multi-Agent Systems, Tool-Augmented LLMs}

\maketitle

\section{Introduction}
\subsection{The Advancement of Agentic LLM Systems}

Large language models (LLMs) are in the midst of a paradigm shift. They are moving from passive, interactive systems to autonomous, goal-directed systems with external actuation capabilities, capable of functioning even in complex settings \cite{dornaika2025agentic}. Initially, large language models (LLMs) were used for one-shot or limited context conversations at best. The models could only generate textual outputs without taking any actions outside their system, could not retain information from previous interactions, and had no tasks left to perform in the future \cite{dimaggio2025toward}.

In contrast to earlier systems, the latest agentic LLMs tend to support memory planning and tool-use plans for achieving self-guided objectives that lie far in the future \cite{janjusevic2025hiding}. These architecturally significant advancements have made LLMs a centerpiece of iterative decision-making processes \cite{xia2025agent0}. Today, agentic models of LLMs can be commanded to break down high-level objectives into intermediate sub-goals, support multistep reasoning, use external "tools" or APIs, and monitor their own internal states based on observations of external outcomes \cite{dimaggio2025toward}. AutoGPT, React-based agents, or tool-based LLM models are examples of systems that better represent the development of language models from passive text-generating machines to generalized computational system controllers [\cite{feng2025emerged, ren2025gtm}]. 

As a result, the scope and operation of large language model-based systems have grown significantly to support increasingly complex and operationally relevant applications \cite{xu2024theagentcompany}, where empirical studies indicate that agent actions can exacerbate model failures and lead to detrimental outcomes \cite{bandi2025rise}, including the generation of incorrect information and irreversible real-world actions such as faulty API calls or unintended code execution \cite{li2025security}, thereby introducing additional operational and security risks. 

\subsection{Security Threats Unique to Agentic Systems}

The security context of agentic LLMs is not comparable to what we are concerned about with traditional LLMs. The major concerns with LLMs, as discussed in most literature, include prompt injection, jailbreaking, data leakage, and misuse, especially during chat-based interactions  \parencite{shenao_wang_34903903}. Although these are valid security concerns, they do not fully capture the security landscape when agentic LLMs are considered.

The security context with agentic LLMs is more alarming because they can interact with other tools, function independently, create their own environment, access memory, and function with other agents \parencite{alsharif_abuadbba_0776f6f8}. When agents are used, these agents can access privileged APIs and change external facts. The agents, especially the retrieval agents, are bound by their external knowledge bases  \parencite{arthur_caetano_187e451f}. When there is a multi-agent system, there is a new channel of attack, especially concerning message injection and impersonation attacks  \parencite{yuntao_wang_bc05705f}. The memory store can allow agents to create channels of attack, especially concerning poisoning and the gradual corruption of the LLM's reasoning and learning processes  \parencite{hanrong_zhang_3eab6154}. Among these, cascading failures represent a particularly concerning threat class. The agents function independently and recursively. This means that if there is an attack, it can cascade through all agents without anyone's knowledge, let alone anyone's control  \parencite{miko_aj_kniejski_5c5c7b36}.

Any part of the agentic LLM setup that can be breached by an attacker and the harmful consequences that follow will affect the entire agentic LLM setup. This suggests that prompt injection has the potential to change an agent's current internal goals, which in turn can influence the planning process, tools, and techniques used  \parencite{sizhe_chen_4ae2f113}. The effects of this then continue over time, becoming harder to identify and address, with the effects not being noticeable until many intermediary actions have been carried out. Even though security issues with agentic LLMs are significant, existing security defenses are based on the era of traditional LLMs. There is, therefore, a growing gap in the capabilities of agentic LLMs and existing security defenses  \parencite{he_feng_b17ea19f}.

\subsection{Scope and Definitions}

In this work, we define agentic LLMs as systems where the primary reasoning element is a large language model (LLM), coupled with an autonomous agent capable of: (1) perceiving and processing inputs from surrounding environments, such as user instructions or external data; (2) performing multi-step planning executing sequences of actions; (3) invoking external tools, APIs, and code execution; and (4) updating its state or memory based on the outcome of previously executed actions.

It is also important to clarify what is not included in the definition of agentic LLMs. These include  purely conversational LLMs, traditional AI agent systems which are not based on LLMs, and studies on LLMs that do not address security or safety aspects related to agentic behaviour. This literature considered in this study spans January 2023 through December 2025. 
This period is crucial since it marks the emergence of agentic LLMs in the literature, focusing on this timeframe enables us to identify vulnerabilities that arise specifically from the agentic properties of LLM-based systems, rather than vulnerabilities associated with LLMs in general.

\subsection{Research Questions}
This research is organized around four research questions that provide a foundation for evaluating the security of agentic LLMs.

\begin{itemize}
    \item \textbf{RQ1:} What types of vulnerabilities arise in agentic LLMs, and at which architectural layers do these vulnerabilities occur?
    \item \textbf{RQ2:} What techniques have been proposed to detect and identify these vulnerabilities, and how effective are they?
    \item \textbf{RQ3:} What mitigation and defense mechanisms have been developed to address vulnerabilities in agentic LLM systems?
    \item \textbf{RQ4:} What critical research gaps remain, and what research areas require urgent investigation?
\end{itemize}
Taken together, these questions provide a framework for the literature review and guide the development of a vulnerability taxonomy and research agenda addressing the security challenges of agentic LLMs.

\subsection{Contributions}

This research is significant for several reasons. To the best of our knowledge, this is the first PRISMA-based systematic review specifically focused on identifying vulnerabilities associated with agentic LLMs. We undertook an extensive review of the literature, in line with the PRISMA guidelines for the year 2020, for 85 peer-reviewed articles (from an initial set of 743) published between 2023 and 2025 in six prominent research databases.

\textbf{Taxonomy of Vulnerability}: Four-layer structure based on agent architecture. We have mapped the vulnerabilities to the different layers of the architecture of the agent, namely perception, brain, action, and interaction.

\textbf{Quantitative Synthesis}: Identification of imbalance and gaps in the literature. Our quantitative synthesis indicates an imbalance between the number of papers focused on attacks (3.9:1) and those focused on defenses, while also identifying significant gaps, such as the lack of coverage for code execution vulnerabilities and security for embodied agents.

\textbf{Research Roadmap}: A comprehensive framework for the research, identifying seven open problems for improving the security of agentic LLM systems.

\subsection{Organization of the Paper}

This paper progresses from establishing context and scope (Section 2) and methodology (Section 3), through background (Section 4), vulnerability analysis (Section 5), and detection/mitigation (Sections 6--7), to identifying research gaps (Section 8) and concluding with key findings and future directions (Section 9).

\section{Related Work and Positioning}

\subsection{Evolution of LLM Security Research}

In the last few years, the pace of research related to the safety and security of large language models has accelerated significantly, reflecting the changing nature and applications of large language models \cite{zangana2024llm}. Traditionally, the research was focused only on the risks associated with large language models, but recently, the academic interest has moved towards the agential behavior of large language models, which includes autonomous task accomplishment, tool use, and long-horizon execution \cite{zhang2024breaking}.

The first set of security research for large language models (LLMs) was initiated with the development and proliferation of large-scale, pre-trained language models \cite{chhabra__anshuman_9d1541d8}. Over a very short period, numerous surveys and empirical studies have identified the security issues faced by conversational and task-oriented large language models, such as prompt injection, jailbreaking, data leakage, memorization, and model abuse \cite{sara_abdali_ce0efbe9}. However, all such works have considered large language models to be passive entities, responding to user queries, without any state, tool, or time-dependent program flow. Nevertheless, the majority of the works have focused only on the surface-level security issues faced by large language models, without considering the security aspects of large language models used in the core decision-making process.

Since the beginning of 2023, the security research for large language model agents has gained momentum, considering the fact that large language model agents have moved beyond the proof-of-concept stage and are being used for practical applications, such as AutoGPT, ReAct, tool-aided large language model agents, and multi-agent systems, which have introduced new security issues, such as uncoordinated failure, tool misuse, environment modification, and multi-step planning challenges \cite{christian_schroeder_de_witt_a4a27db7}.

\sloppy

\subsection{Existing Surveys and Their Limitations}
\begin{figure}[t]
    \centering
    \includegraphics[width=0.65\linewidth]{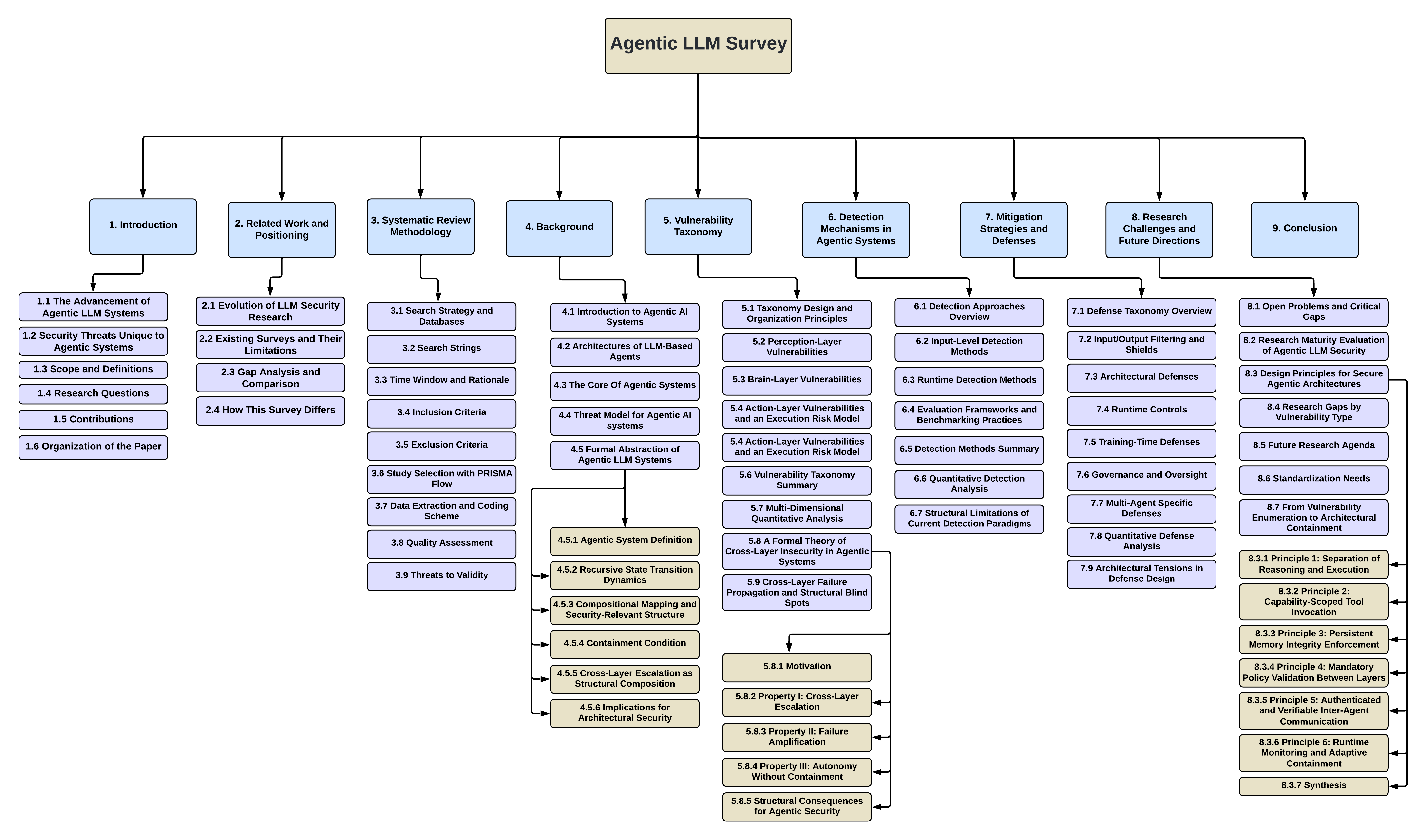} 
    \caption{Organization of the Paper}
    \label{fig:organization}
\end{figure}

Existing surveys address related issues in LLM or AI security, but none provides a dedicated architectural vulnerability taxonomy for agentic LLM systems grounded in a systematic review of the literature.

During the pre-LLMs period, security surveys focused on software agents, multi-agent systems, and robotic agents \cite{yuntao_wang_8d255dc9}. Although these studies address security issues such as trust, coordination problems, and adversarial situations, their results apply only to symbolic agents. As a result, they do not address potential security concerns related to LLM systems, such as prompt injection, emergent reasoning failure, or tool misuse due to language.

Some recent surveys have focused on agentic AI and autonomous LLMs. However, most existing reviews do not report search strategies, inclusion and exclusion criteria, or assess the completeness of the provided information. The majority of existing surveys are narrative and emphasize security mechanisms in association with system challenges or governance. Narrative reviews in this space tend to address security mechanisms at a surface level; while some address multi-agent systems and tool-based security, none adequately covers security vulnerabilities related to agent architecture.

Although existing surveys cover LLM safety, adversarial machine learning, and AI agent security, they mostly focus on model-level attacks and do not provide an architectural view on agential LLM vulnerabilities. This survey is unique in that it (1) is underpinned by a systematic review methodology based on PRISMA, (2) offers a four-layer architectural taxonomy for LLM vulnerabilities, and (3) provides a quantitative gap analysis on security research in the architectural layers.

\subsection{Gap Analysis and Comparison}

This study presents a systematic literature review on security vulnerabilities of agentic large language model systems by following a PRISMA protocol, which synthesizes and organizes existing literature on different components.

We provide an abridged overview of representative historical survey studies on several key dimensions, as shown in Table 1. The survey covers all aspects of agent architectures, ranging from single-agent and multi-agent systems to tool-augmented and RAG-based systems, as well as security requirements, including vulnerabilities, detections, and mitigations. The following section presents three major gaps found in previous studies. The gaps include: (1) agent architectures, (2) focus on weaknesses, not detection and mitigation, and (3) lack of formalism found in survey studies.

\begin{table*}[h!]
\caption{Coverage Comparison with Previous Surveys. 
\textbf{Legend:} \cmark{} = In-depth coverage / Yes, 
\pmark{} = Partially covered, 
\xmark{} = Not covered.}
\centering
\resizebox{\textwidth}{!}{%
\begin{tabular}{l c c c c c c c c c}
\toprule
\textbf{Reference} & \textbf{Year} & \textbf{PRISMA} & \textbf{Arch. Taxonomy} & \textbf{Single Agent} & \textbf{Multi-Agent} & \textbf{Tool-using} & \textbf{Vuln. Categories} & \textbf{Detection} & \textbf{Mitigation} \\
\midrule

\cite{li2025security} & 2025 & \xmark & \xmark & \pmark & \xmark & \xmark & \cmark & \xmark & \cmark \\
\cite{shayegani2023survey} & 2023 & \xmark & \xmark & \pmark & \cmark & \xmark & \cmark & \xmark & \cmark \\
\cite{brohi2025research} & 2025 & \xmark & \xmark & \pmark & \cmark & \cmark & \xmark & \cmark & \xmark \\
\cite{wang2025unique} & 2024 & \xmark & \xmark & \pmark & \xmark & \xmark & \cmark & \xmark & \cmark \\
\cite{raza2025trism} & 2025 & \xmark & \pmark & \pmark & \cmark & \cmark & \cmark & \xmark & \xmark \\
\cite{deng2025ai} & 2024 & \xmark & \pmark & \pmark & \xmark & \xmark & \cmark & \xmark & \cmark \\
\cite{gan2024navigating} & 2024 & \xmark & \xmark & \pmark & \xmark & \xmark & \cmark & \cmark & \cmark \\
\cite{feng2025emerged} & 2024 & \xmark & \xmark & \pmark & \xmark & \xmark & \cmark & \xmark & \xmark \\
\cite{zhu2025master} & 2025 & \xmark & \xmark & \pmark & \cmark & \xmark & \cmark & \xmark & \cmark \\
\cite{wang2025unique} & 2025 & \xmark & \xmark & \pmark & \xmark & \xmark & \cmark & \xmark & \cmark \\
\cite{ma2025safety} & 2025 & \xmark & \pmark & \pmark & \xmark & \xmark & \cmark & \cmark & \cmark \\
\cite{xu2025llmcyber} & 2025 & \pmark & \xmark & \pmark & \xmark & \xmark & \cmark & \cmark & \cmark \\
\cite{shi2024llmsafety} & 2024 & \xmark & \xmark & \pmark & \xmark & \xmark & \cmark & \xmark & \cmark \\
\cite{ma2025safety} & 2025 & \xmark & \pmark & \pmark & \xmark & \xmark & \cmark & \cmark & \cmark \\
\cite{choi2025review} & 2025 & \xmark & \xmark & \pmark & \cmark & \xmark & \cmark & \pmark & \cmark \\
\cite{yang2024security} & 2024 & \xmark & \pmark & \pmark & \cmark & \xmark & \cmark & \xmark & \cmark \\
\cite{barua2024exploring} & 2024 & \xmark & \xmark & \pmark & \xmark & \cmark & \xmark & \xmark & \xmark \\

\midrule
\textbf{This Survey (Ours)} & 2025 & \cmark & \cmark & \cmark & \cmark & \cmark & \cmark & \cmark & \cmark \\

\bottomrule
\end{tabular}%
}
\end{table*}

\paragraph{PRISMA-based Research Methodology:} This is the first PRISMA 2020-compliant systematic review in agentic LLM security, with Cohen's $\kappa$ values of 0.88 and 0.93, unlike prior surveys that omit search strings and inter-rater statistics.

\paragraph{Quantitative Analysis and Gap Identification:} Our quantitative approach surfaces a 14-fold disparity between perception-layer (65.9\%) and action-layer (4.7\%) coverage, and the critical underrepresentation of code-execution agents (3.5\%), which narrative surveys cannot detect.

\paragraph{Architecture-to-Vulnerability-to-Defense Approach:} By linking architectural components to their vulnerabilities and defenses, this framework provides a component-based basis for designing agent security.

Although 85 papers may appear modest relative to broader LLM safety surveys, this reflects the emergent nature of agentic LLM security as a distinct subfield.

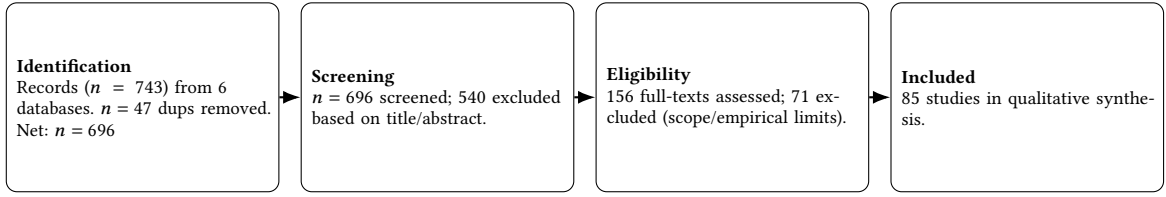
\begin{figure}[t]
  \centering
  \footnotesize
  \begin{tikzpicture}[
    box/.style={
      draw, rounded corners, align=left, 
      text width=0.22\textwidth, 
      inner sep=4pt,
      minimum height=2.5cm 
    },
    arr/.style={-Latex, thick}
  ]

    \node[box] (id) {\textbf{Identification}\\ Records ($n=743$) from 6 databases. $n=47$ dups removed. Net: \textbf{$n=696$}};

    \node[box, right=8pt of id] (scr) {\textbf{Screening}\\ $n=696$ screened; $540$ excluded based on title/abstract.};

    \node[box, right=8pt of scr] (elig) {\textbf{Eligibility}\\ $156$ full-texts assessed; $71$ excluded (scope/empirical limits).};

    \node[box, right=8pt of elig] (inc) {\textbf{Included}\\ $85$ studies in qualitative synthesis.};

    \draw[arr] (id) -- (scr);
    \draw[arr] (scr) -- (elig);
    \draw[arr] (elig) -- (inc);

  \end{tikzpicture}
  \caption{Condensed PRISMA flow diagram of the study selection process.}
  \label{fig:prisma_horiz}
\end{figure}

\subsection{Search Strategy, Search Strings and Databases}

We searched six major databases, which cover the areas of computer security, artificial intelligence/machine learning, and systems research: IEEE Xplore, ACM Digital Library, arXiv, Scopus, Web of Science, and Google Scholar. These databases collectively cover over one million publications in computer science and include the primary venues for agentic LLM security research. We used multiple strategies: Boolean search queries based on carefully selected keywords, snowball sampling based on the initial set of seed papers, and examining existing conference proceedings from top venues such as NeurIPS, ICML, ICLR, USENIX Security, ACM CCS, and IEEE S\&P.

We constructed a search strategy using three complementary query groups combined as: (Group 1) AND (Group 2) NOT (Group 3). Group 1 captures agent-related terms (e.g., agentic LLMs, autonomous agents, tool-augmented systems), Group 2 targets security concepts (e.g., attacks, vulnerabilities, prompt injection, mitigation), and Group 3 excludes out-of-scope domains (e.g., traditional, reinforcement learning, or game agents). Queries were adapted to the syntax of databases including IEEE Xplore, ACM Digital Library, Scopus, Web of Science, arXiv, and Google Scholar.

\subsection{Data Extraction and Coding Scheme}

We created a structured data extraction tool to collect relevant data from each of the 85 papers. We extracted identifiers (title, authors, year, venue), agent type classification, primary type of vulnerability and architectural layer, threat model, nature of contribution (attack, defense, framework, or benchmark), experimental setup, and key findings including limitations identified by the authors.

To assess the reliability of the coding process, a random stratified sample of 15 papers (17\% of the corpus) was independently coded by a second reviewer. Inter-rater agreement for vulnerability classification was 94\%, and agreement for paper categorization (attack, defense, framework) was 98\%. Cohen's $\kappa$ was computed to account for chance agreement, yielding $\kappa = 0.88$ for vulnerability classification and $\kappa = 0.93$ for paper categorization, indicating near-perfect agreement. Discrepancies were resolved through discussion and consensus, and the taxonomy definitions were refined accordingly.

\subsection{Quality Assessment}

\begin{figure}[!htbp]
  \centering
  \includegraphics[width=0.5\linewidth]{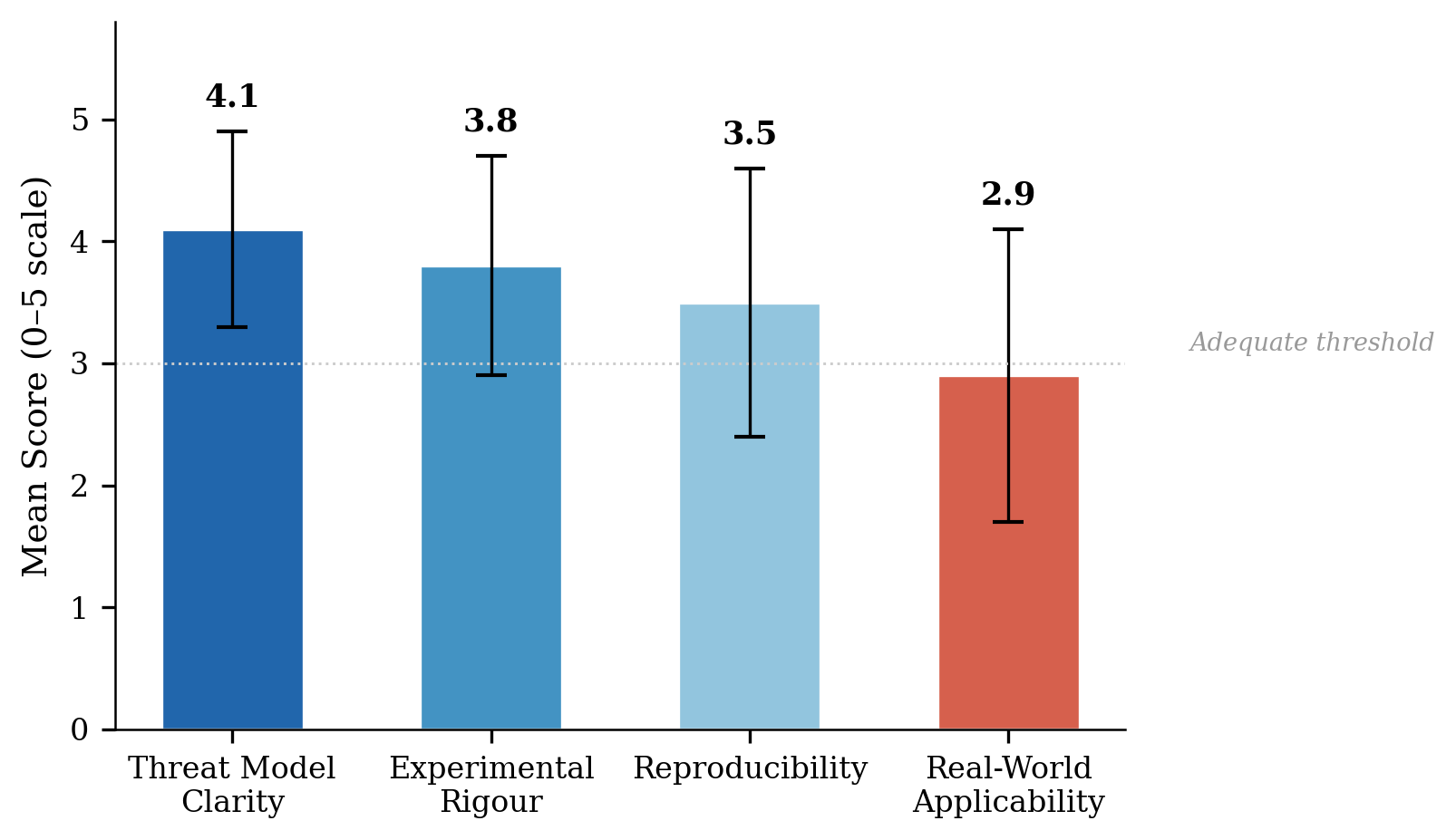}
  \caption{Overall quality assessment distribution of the 85 included papers.}
  \label{fig:quality_assessment}
\end{figure}

Quality was assessed across four dimensions (0--5 scale): Threat Model Clarity (mean 4.1, $\sigma=0.8$), Experimental Rigour (mean 3.8, $\sigma=0.9$), Reproducibility (mean 3.5, $\sigma=1.1$), and Real-World Applicability (mean 2.9, $\sigma=1.2$); 29.9\% rated excellent, 40.2\% good, 21.8\% fair, and 8\% preliminary (Figure~\ref{fig:quality_assessment}).

\subsection{Threats to Validity}

We identify several potential threats to the validity of this study:

\textbf{Selection Bias:} Although the search string used in this research inquiry is comprehensive, some publications, especially those not written in the English language or those which use unconventional terms, might not be found in the results obtained from the search string used in the research inquiry. Furthermore, it is also possible that some publications received in 2025 might not be included in the databases used for the research inquiry, and this might be addressed by the use of a snowball sampling technique.

\textbf{Publication Selection Bias:} The literature suggests a higher probability of publication for research that reports positive findings compared to null findings or unsuccessful approaches. This could result in an overestimation of the effectiveness reported in works on mitigation and detection.

\textbf{Reliability of Coding:} In view of resource availability, it was mainly the primary reviewer who coded the majority of papers included in the corpus. To assess reliability, 17\% of the corpus, i.e., 15 papers, was chosen through stratified sampling. These papers were then independently coded by a second reviewer. There was substantial agreement between two coders, as reflected by high values of kappa statistics, 0.88 and 0.93, for vulnerability classification and paper classification into attack, defense, and framework, respectively.

\textbf{Fast-Moving Area:} The field is in a state of rapid development. Publications released from Q4 2024 onwards may not have been adequately indexed. Any publications released after December 2025 are outside the scope of this review. Perhaps a living review would be more appropriate in the future.

\section{BACKGROUND}

\subsection{Introduction to Agentic AI Systems}

Agentic Large Language Model (LLM) systems represent a remarkable leap forward in the evolution of artificial intelligence, marking the shift from passive query-and-response to autonomous or semi-autonomous intelligent agents with goals and objectives \cite{ali__mohamad_abou_8a36d30e}. The development of agentic artificial intelligence, including LLMs within a larger architecture, overcomes these challenges. The current era of agentic LLMs is based on a series of critical technological advancements. These systems rely on four enabling capabilities: in-context learning for short-term memory and adaptive reasoning; structured tool and function calling for interaction with external systems; retrieval-augmented generation (RAG) for dynamic knowledge access; and persistent memory for retaining state across episodes~\cite{molinari2025pervasive, gan2025ragmcp, singh2025agentic, wu2025memory}.

One of the first agent-based system frameworks developed was ReAct. It was possible to obtain task performance, interpretability, and stability \cite{yao2022react}. While the decomposition of tasks into sequences of thought-action observation cycles was key to making agentic LLMs practical, ReAct provided a framework that served as an engineering blueprint for constructing such systems and led to rapid uptake in both research prototypes and production systems \cite{yao2022react}. Since then, they have been used in various applications, such as intelligent coding agents, automated workflows, research bots, customer support orchestration, and decision support itself \cite{wang2025evoagentx}. 

\subsection{Architectures of LLM-Based Agents}

The literature survey discussed here covers a wide range of architectures of agent system designs, reflecting various compromises with respect to important design goals such as autonomy, scalability, coordination, and safety. The models discussed can be broadly grouped into seven categories.

\textbf{Single-agent systems} Single-agent systems consist of one LLM instance that handles all reasoning, decision making, and execution towards a goal \cite{masterman2024landscape}. These systems are conceptually straightforward, and their dynamic characteristics have been extensively explored in personal assistants, task automation agents, and research platforms, including LangeChain-based agents \cite{barua2024exploring}. On the other hand, single-agent systems are simpler to design and reason about but can face challenges in completing complex tasks that involve parallel reasoning, specialization, or adversarial assessment \cite{kim2025scaling}.

\textbf{Multi-agent systems} Agent systems consist of several LLM instances communicating among themselves. Such agents may cooperate toward joint goals, discuss possible alternative solutions to a problem, or organize themselves into hierarchies in which higher-level agents decompose tasks into subtasks and distribute them to specialized subagents. Multi-agent systems have been investigated for collaborative problem-solving, agent debate for improving reasoning accuracy, and simulation of social or economic environments \cite{han2024llm}. However, they also introduce novel coordination and attack-surface issues related to inter-agent communication.

\textbf{Tool-Augmented Agents} By contrast, tool-augmented agents extend LLMs with the possibility of calling external tools and APIs \cite{zhang2024whenmeet}. Such tools may include, for example, programming language interpreters, database query engines, mathematical solvers, or enterprise software application interfaces. 

\textbf{Retrieval-augmented generation (RAG) agents} RAG agents incorporate retrieval mechanisms inside the agent loop. These agents can retrieve the latest and/or domain-specific information by querying vector databases, document collections, or knowledge graphs \cite{xu2024genai}. 

\textbf{Code execution agents} Code execution agents can be used to create and execute code, in secure (e.g., sandboxed) or untrusted operating environments or in a trusted executing environment \cite{mo2024trembling}. These agents are well-suited for data analysis, software development and system management. However, they are among the highest-risk categories, as bugs in code generation or execution can result in arbitrary code execution, data leakage, or system compromise \cite{navneet2025rethinking}.

\textbf{Web-crawling agents} Web-crawling agents can be capable of traversing the web and following links, understanding retrieved pages' content, and engaging in navigation decision-making. These agents are well-suited for scraping, monitoring, and intelligence-gathering, but are subject to malicious content injection and fraudulent interface attacks \cite{zhang2025deepresearch}.
\textbf{Embodied agents} Embodied agents function within a real or virtual world, and perform the integration of perceptual sources (e.g., visual information or sensor data) with action and planning \cite{wu2024dissecting}. Though rare in our corpus, embodied agents represent high-stakes deployment settings where physical safety and robustness are paramount \cite{corban_g__rivera_e45ec196}.

\subsection{The Core Of Agentic Systems}
While agent architectures can differ, many agentic AI systems share four primary layers \cite{ibrahim_adabara_885ff5cf}. These layers provide a security-related abstraction that can be used to model system behavior, organize the assessment of system vulnerabilities, and identify containment boundaries in which policy enforcement should take place \cite{david_shapiro_bc4eca02}.

\textbf{The Perception Layer}\\
The perception layer is responsible for receiving and interpreting inputs from users, external systems, and environmental sources. It is in charge of consuming and interpreting inputs from a variety of sources: user commands, retrieved documents, tool responses, sensor readings and inter-agent messages \cite{zahid2025enhancing}. This layer converts raw, typically ill-structured data to reasoning-friendly representations \cite{sager2025comprehensive}. The perception layer is most vulnerable to attacks such as prompt injection, jailbreak attempts, and adversarial input manipulation, as it processes inputs from external sources \cite{deng2025ai}. Implementation components include: user input interfaces, retrieval pipelines (RAG),input preprocessing modules, external data sources

\textbf{The Brain Layer}\\
The brain layer performs reasoning, planning, and decision-making based on perceived inputs and memory state. This layer is the cognitive centre of the agent. It consists of the LLM, as well as connected planning and reasoning, goal maintenance, and memory systems \cite{dong2024jailbreaking}. This level describes what the agent wants to do, and why. At this level, vulnerabilities may include manipulation of reasoning, behaviour insertion of backdoor behaviours into model weights, and goal-hijacking attacks that redirect agent objectives to an unintended target. Implementation components include LLM reasoning modules, planning and task decomposition mechanisms, chain-of-thought reasoning frameworks, and goal and state management components.

\textbf{The Action Layer}\\
The action layer executes decisions generated by the reasoning component by invoking tools, APIs, or external systems. This layer executes the decision made by the brain layer \cite{liu2025advances}. These actions might involve invoking tools, running code, interacting with APIs, or controlling other physical systems. This layer is particularly fragile since every action has clear side effects. Some attacks, such as tool misuse, function or code injection, and hijacking, can have devastating effects, including unwanted system alterations and information leakage \cite{noman2023code}. Implementation components include tool invocation frameworks, API execution modules, code execution environments, and automation scripts.

\textbf{The Interaction Layer}\\
The interaction layer manages communication between agents, external systems, and shared environments. This layer is responsible for agent-environment communication, including other agents(s), memory store(s) and external service(s) \cite{he2024security}. This layer controls the behaviour of information entering and exiting the system. There are vulnerabilities here, such as message injection, memory poisoning, and manipulation of shared environmental state, which can have cascading effects across agent behaviours \cite{he2025comprehensive}.Through such an organisation of agentic systems, it is possible to systematically analyse and categorise security threats and design the most suitable countermeasures. Implementation components include multi-agent communication protocols, message passing systems, environment interfaces, and shared memory channels.

\begin{table}[h]
\centering
\caption{Mapping of Existing Agent Framework Components to the Proposed Four-Layer Taxonomy}

\begin{tabular}{l l}
\toprule
\textbf{Existing Framework Component} & \textbf{This Survey Layer Mapping} \\
\midrule
BDI: Beliefs & Perception / Memory Layer \\
BDI: Desires / Intentions & Brain Layer \\
ReAct: Observation & Perception Layer \\
ReAct: Reasoning / Thought & Brain Layer \\
ReAct: Action & Action Layer \\
AutoGPT: Tool Invocation & Action Layer \\
Multi-Agent Communication & Interaction Layer \\
\bottomrule
\end{tabular}
\end{table}

\subsection{Threat Model for Agentic AI systems}

In order to assess the security of agentic systems, we follow a threat model, which distinguishes between the type and extent of knowledge and access available to an attacker \cite{christodorescu2025systems}. This is a widely used taxonomy in security research, offering a theoretical framework for evaluating the severity of security flaws.

\textbf{White-box Attack Model:} 

In this approach, it is assumed that the attacker can access not only the model's weights, architectures, and training information, but also all of the model's operational parameters \cite{tianyang_wang_dcabf862}. This is the most extreme approach and is used in open-source, insider threat, and model leakage risk scenarios. In this white-box approach, a new type of attack is introduced: targeted attacks, such as backdoors in the model's weights, and precise attacks, such as manipulating reasoning \cite{yuxi_li_4a4097e9}.

\textbf{Black-box Attack Model:} 

In the black-box threat model, it is assumed that the attacker can only access the model's inputs and outputs, typically via an API or UI. This is the most realistic scenario for commercial applications. Even in such an unfavourable scenario, it is possible to carry out proactive attacks, such as early injections and jailbreaking, that will indirectly compromise the agent's behaviour.

\textbf{Grey-box Attack Model:} The threat model for this framework is grey-box and thus falls between these two ends of the spectrum. Adversaries have partial information (e.g., architectural designs, training regimes or access to logs and documentation), but not all model parameters. These may correspond to grey-box settings such as collaborative developmentsettings, supply chain environments, or compromised insiders with only limited authority \cite{mazzone2024privacy}.

For all the vulnerability classes explored in this study, we estimate the base attacker skill level required for successful exploitation. Interestingly, several high-impact attacks on agentic systems (in particular, prompt injection and tool abuse) are solvable in a black-box setting. More advanced attacks, such as persistent backdoors or targeted reasoning corruption, can often only be mounted by attackers with grey-box or white-box access \cite{liu2025advances}.

\section{Vulnerability Taxonomy}
This section introduces a taxonomy of vulnerabilities in agentic LLM systems, grouped according to their occurrence in different architectural layers.

\subsection{Taxonomy Design and Organization Principles}

The 13 vulnerability types were identified through an inductive open-coding process applied to the 85 papers in the corpus during the data extraction stage. Each paper was coded according to the primary attack mechanism described, and the resulting codes were iteratively consolidated and refined until a stable set of categories emerged. While the four-layer architecture provided the organizational framework, the vulnerability types within each layer were not predefined. Through three rounds of refinement, an initial set of 23 candidate categories was consolidated into the final 13 by merging closely related attack mechanisms. While vulnerabilities are classified according to their primary architectural layer, several attacks may propagate across layers during execution. Such cross-layer effects are indicated explicitly in the taxonomy tables.

Our vulnerability taxonomy categorizes 13 vulnerability types across four architectural layers. Rather than organizing by attack name (which is used inconsistently across the literature), we categorize vulnerabilities by the system component in which they arise. This component-based approach provides actionable insights; i.e., practitioners responsible for designing each component can focus on the vulnerabilities most relevant to that component of the system. The taxonomy is defined by two dimensions: (1) architectural layer, which describes the location of the vulnerability, and (2) vulnerability type, which describes the attack mechanism. The two-dimensional taxonomy enables an understanding of beyond merely the scope of analysis in terms of the number of vulnerabilities analyzed, but also the level of analysis in terms of the amount of effort put into each layer. The four layers are mutually exclusive with regard to primary function: every architectural component is assigned to exactly one layer depending on the primary function that dominates in the perception-execution pipeline, even though some implementation details may occasionally span multiple layers.

\subsection{Perception-Layer Vulnerabilities}

Perception-layer vulnerabilities occur during input processing, prior to the agent's Reasoning system kicking in. This category is the most extensively researched in our corpus, appearing in 56 of 85 papers (66\%). Direct 

\textbf{Prompt Injection:} When an attacker gains direct control over part of the prompt \cite{xiaogeng_liu_df79f45f}. We demonstrate the basic attack of instruction overwriting: the attacker-supplied content changes what the agent is supposed to maximize in reward, and the instruction not to follow previously applied constraints. 
In contrast to the manipulation of the process of reasoning, which activates the cognitive chain at the brain layer, prompt injection affects the perception layer by modifying the inputs given to the agent before the process of reasoning starts.

\textbf{Indirect Prompt Injection:} Malicious instructions are placed in input sources and then processed later by the agent \cite{chongyang_shi_062fefe7}. Examples are web pages loaded by the agent being poisoned, emails read by the agent being malicious, or database records retrieved being corrupted \cite{zehang_deng_1ca5d202}. This class is found in the same research papers as direct injection, \cite{sippo_rossi_f6494e89}. 

\textbf{Jailbreaking Attacks:} Try to bypass the agent's safety barrier with a well-crafted prompt. Jailbreaks do not directly inject instructions; rather, they exploit the model's tendency to reproduce patterns from the context provided by the prompt \cite{benji_peng_910433b5}.  

\textbf{Adversarial Input Perturbations:} Adversarial samples, perturbation attacks and input transformations that have been constructed to be misclassified \cite{sunwoo_lee_380565e9}. They differ from prompt injections in that their structure is less blunt and does not require text-based prompts. 

The layer indicated in the following table represents the primary location where the vulnerability arises, although certain attacks may propagate to other layers during system operation.

\begin{table*}[!htbp]
\centering
\caption{Vulnerability Taxonomy by Agent Architecture. Reference numbers in brackets indicate studies that have examined each vulnerability type within the corresponding agent architecture. An em dash (\textemdash) denotes no identified studies for that combination.}
\label{tab:vuln_taxonomy}
\renewcommand{\arraystretch}{1.3}
\small
\begin{tabular}{@{}p{3.2cm}cccccccc@{}}
\toprule
\textbf{Vulnerability Type} & \textbf{Single Agent} & \textbf{Multi-Agent} & \textbf{Tool-Aug.} & \textbf{RAG-based} & \textbf{Web Agent} & \textbf{Code-Exec} & \textbf{Embodied} \\
\midrule

\multicolumn{8}{l}{\textbf{PERCEPTION LAYER}} \\
\addlinespace[2pt]

Direct Prompt Injection 
  & \cite{han2024llm} 
  & \cite{pedro2023prompt, fang2025breaking, cui2025mad} 
  & \cite{zhan2024injecagent, debenedetti2024agentdojo} 
  & \cite{li2024targeting} 
  & \cite{li2025security, wu2024wipi} 
  & \cite{dornaika2025agentic} 
  & \textemdash \\

Indirect Prompt Injection 
  & \textemdash 
  & \cite{datta2025agentic, deng2025ai} 
  & \cite{zhan2024injecagent, ferrag2025prompt} 
  & \cite{dong2024jailbreaking, li2024targeting, ju2024flooding} 
  & \cite{li2025security, wu2024wipi} 
  & \cite{deng2025ai} 
  & \textemdash \\

Jailbreaking Attacks 
  & \cite{christodorescu2025systems, he2025comprehensive} 
  & \cite{fang2025breaking, gan2024navigating, he2024security, raza2025trism, zhou2024defending} 
  & \textemdash 
  & \cite{ghosh2025agentic, feng2025emerged, yu2024llmvirus, choi2025review} 
  & \textemdash 
  & \textemdash 
  & \textemdash \\

Adversarial Perturbations 
  & \cite{shayegani2023survey} 
  & \cite{evertz2024whispers, gan2024navigating, jiang2025safety, narajala2025securing, pan2025measuring} 
  & \cite{jiang2025safety} 
  & \cite{liu2025advances, dimaggio2025toward} 
  & \cite{jiang2025safety, li2025mind} 
  & \textemdash 
  & \cite{evertz2024whispers, liu2024compromising} \\

\addlinespace[4pt]

\multicolumn{8}{l}{\textbf{BRAIN LAYER}} \\
\addlinespace[2pt]

Backdoor Attacks 
  & \textemdash 
  & \cite{he2024words, singh2025agentic} 
  & \textemdash 
  & \cite{barua2024exploring, dong2024jailbreaking} 
  & \textemdash 
  & \textemdash 
  & \cite{liu2024compromising} \\

Reasoning Manipulation 
  & \textemdash 
  & \cite{gan2024navigating, mo2024trembling} 
  & \textemdash 
  & \cite{evertz2024whispers} 
  & \textemdash 
  & \textemdash 
  & \cite{evertz2024whispers} \\

Goal/Plan Hijacking 
  & \textemdash 
  & \cite{janjusevic2025hiding, ren2025gtm} 
  & \cite{lei_wang_f6fad767} 
  & \textemdash 
  & \cite{li2025mind} 
  & \textemdash 
  & \textemdash \\
 
Memory Poisoning 
  & \textemdash 
  & \textemdash 
  & \textemdash 
  & \cite{ju2024flooding} 
  & \textemdash 
  & \textemdash 
  & \textemdash \\

\addlinespace[4pt]

\multicolumn{8}{l}{\textbf{ACTION LAYER}} \\
\addlinespace[2pt]

Tool Manipulation 
  & \textemdash 
  & \textemdash 
  & \cite{barua2024exploring, zhan2024injecagent, debenedetti2024agentdojo, ferrag2025prompt} 
  & \textemdash 
  & \textemdash 
  & \textemdash 
  & \textemdash \\

Function Hijacking 
  & \textemdash 
  & \textemdash 
  & \cite{janjusevic2025hiding, ferrag2025prompt} 
  & \textemdash 
  & \textemdash 
  & \textemdash 
  & \textemdash \\

Code Injection 
  & \textemdash 
  & \textemdash 
  & \textemdash 
  & \textemdash 
  & \textemdash 
  & \cite{deng2025ai, dornaika2025agentic, triedman2025multi} 
  & \textemdash \\

Sandbox Escape 
  & \textemdash 
  & \textemdash 
  & \textemdash 
  & \textemdash 
  & \textemdash 
  & \cite{patlan2025real, triedman2025multi} 
  & \textemdash \\

\addlinespace[4pt]

\multicolumn{8}{l}{\textbf{INTERACTION LAYER}} \\
\addlinespace[2pt]

Inter-Agent Msg. Injection 
  & \textemdash 
  & \cite{pedro2023prompt, datta2025agentic, deng2025ai, gan2024navigating} 
  & \textemdash 
  & \textemdash 
  & \textemdash 
  & \textemdash 
  & \textemdash \\

Agent Impersonation 
  & \textemdash 
  & \cite{narajala2025securing, zhu2025master} 
  & \textemdash 
  & \textemdash 
  & \textemdash 
  & \textemdash 
  & \textemdash \\

Knowledge Base Corruption 
  & \textemdash 
  & \textemdash 
  & \textemdash 
  & \cite{dong2024jailbreaking, li2024targeting, ju2024flooding} 
  & \textemdash 
  & \textemdash 
  & \textemdash \\

\bottomrule
\end{tabular}
\end{table*}

\begin{figure}[!htbp]
  \centering
  \includegraphics[width=0.6\linewidth]{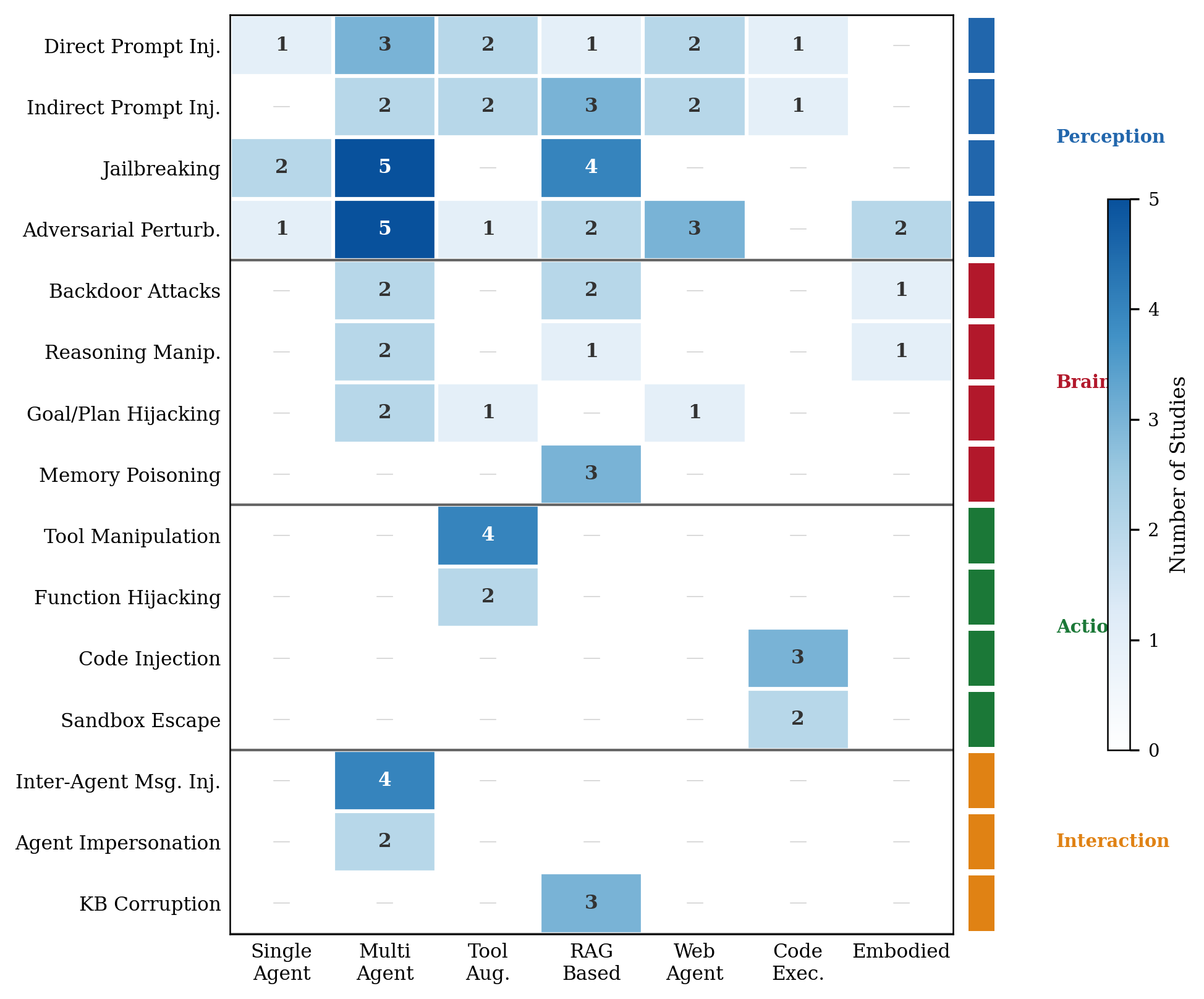}
  \caption{Heatmap of vulnerability distribution across architectural layers and agent types (visual representation of Table 2).}
  \label{fig:vuln_heatmap}
\end{figure}

\subsection{Brain-Layer Vulnerabilities}

Attacking the brain layer focuses on the agent's reasoning, planning, and decision-making. To date, our survey shows that these vulnerabilities are present in 35/85 papers (41\%), indicating that this layer has received significantly less research attention than perception-layer attacks.

\textbf{Backdoor Attacks}: Backdoors are latent triggers introduced at training time that fire only when a specific input is observed \cite{emily_wenger_2581f438}. In an agentic setting, a backdoor code can force the agent to follow rogue guidelines, place unauthorised calls to applications and forge logs daily \cite{l_o_boisvert_a93c47d6}.

\textbf{Manipulation of Reasoning}: This uses the reasoning power of the agent to reach incorrect conclusions \cite{y__huang_d3c42ba1}. That includes introducing subtle errors into the reasoning chain that are individually plausible but collectively lead to incorrect conclusions \cite{rishane_dassanayake_c1036e32}. This differs from prompt injection because the adversarial influence targets the reasoning process itself rather than the agent's input.

\textbf{Goal Hijacking and Action Hijacking}: The agent's goals are changed or the actions, which it intended to perform, are steered towards other than desired directions \cite{yuyou_gan_d2122eda}. Goal hijacking is especially harmful for agent systems, as it may lead the agent to devote its efforts to the attacker's objectives rather than its initial goals \cite{abhinav_rao_9d5fed38}. 

\subsection{Action-Layer Vulnerabilities and an Execution Risk Model}

The execution layer represents a high-risk attack surface in agentic systems, as it interfaces with external tools, APIs, and environments, enabling immediate and potentially irreversible effects. When considering the different layers, perception and reasonings tend to capture the majority of the attention, but the highest-consequence failures arise at the execution layer \cite{xing_wu_8ee09deb}. This is the part of the agentic LLM system that impacts the outside world. Calling tools, calling different APIs, reading and writing files, network communication, and even driving physical actions are all examples of the execution layer \cite{bang_liu_60553ab1}. Mistakes in the execution layer do not just misrepresent facts; they change the state of the world in a very observable and sometimes irreversible fashion \cite{deven_r__desai_4d331163}.

However, despite the high stakes of the execution layer, the majority of studies rarely explore the compromise of the execution layer in depth. This discrepancy can be attributed to the mismatch between the emphasis placed on different components of the LLM system and the practical realities of its deployment \cite{qiang_hu_787458bc}. To further clarify the issue, the insecurities related to the execution layer are discussed from three different perspectives: privilege, external state modification, and the possibility of irreversible actions.

\subsubsection{Tool Manipulation}
Tool manipulation occurs when an attacker forces an agent to invoke 
unintended tools or misuse legitimate ones, potentially triggering 
unauthorized API calls, data exfiltration, or resource abuse~\cite{barua2024exploring, 
zhan2024injecagent}. The relative underrepresentation of this attack 
class in the literature is discussed in Section~5.7.

\subsubsection{Code Injection}
Code injection vulnerabilities occur when agents generate and execute code derived from untrusted input sources. In several agent systems, code produced by the agent is executed directly by an interpreter. Empirical studies show that adversarial prompts or retrieved data can inject malicious instructions into the generated code, enabling arbitrary command execution or unauthorized access to files and networks \cite{bandi2025rise}. This vulnerability class represents one of the most impactful attack types in agentic systems.

\subsubsection{Sandbox Escape}
Sandbox escape vulnerabilities occur when attackers bypass isolation mechanisms intended to restrict the execution of code generated by agents. Although containerization and sandboxing are commonly used to limit code execution privileges, empirical studies show that attackers can exploit misconfigurations or interpreter capabilities to allow malicious code to access host resources outside the restricted environment \cite{debenedetti2024agentdojo}.

\subsection{Interaction-Layer Vulnerabilities}

Interaction-layer vulnerabilities target communication channels between agents and external systems, other agents, and memory stores \cite{daniel_jones_58fcc893}.

\textbf{Agent-to-Environment Attacks}

This includes webpage poisoning, where web content served to web-browsing agents contains malicious instructions, environment manipulation, where the observation space contains injected content, and sensor manipulation for embodied agents \cite{shaked_zychlinski_88a84018}. Our analysis identifies these in 4 papers.

\textbf{Agent-to-Agent Attacks} 

This is specific to multi-agent systems in which agents communicate with each other. Message injection occurs when inter-agent communications are tampered with \cite{yuntao_wang_f3e3bb7f}. In contrast to traditional distributed-system message tampering, which focuses on protocol-level integrity, inter-agent message injection in LLM-based systems manipulates the semantic content of messages, allowing malicious instructions to appear as legitimate tasks because agent communication typically lacks cryptographic message authentication. 

Agent impersonation occurs when one agent spoofs another's identity. Unlike traditional distributed systems where identity is established through cryptographic authentication, agentic systems often rely on conversational context and role descriptions, which can be spoofed through prompt manipulation or message injection.

\textbf{Agent-to-Memory Attacks} 

Include memory poisoning, where the agent's persistent memory is corrupted with false information, and RAG poisoning, where the knowledge base content is contaminated \cite{hanrong_zhang_3eab6154}. Our analysis found this dimension represented in three papers addressing RAG-based agents.

\subsection{Vulnerability Taxonomy Summary}

This section presents an exhaustive synthesis of the identified vulnerabilities. Table 4 reports the frequency of vulnerability occurrences rather than unique studies, as individual papers may contribute to multiple categories (e.g., indirect prompt injection and tool manipulation).

\begin{table*}[t]
  \caption{Component-Level Occurrence of Vulnerability Types Across Architectural Layers. 
  Vulnerabilities are classified by the primary layer where the attack originates.}
  \label{tab:agent_attacks}
  \Description{A comprehensive table of agentic vulnerabilities across Perception, Brain, Action, and Interaction layers, including attack vectors, examples, impact, severity, frequency counts, and corresponding literature citations.}
  
  \small
  \begin{tabularx}{\textwidth}{@{} l l X c c p{4.5cm} @{}}
    \toprule
    \textbf{Component} & \textbf{Attack Vector} & \textbf{Example Attack} & \textbf{Sev.} & \textbf{Cnt.} & \textbf{Key References} \\
    \midrule
    \multicolumn{6}{l}{\textit{PERCEPTION LAYER}} \\
    \midrule
    Input Processing & Direct Prompt Inj. & Malicious user instructions & High & 12 & \cite{pedro2023prompt, zhan2024injecagent, ferrag2025prompt, han2024llm, li2025security, cui2025mad, li2024targeting} \\
    Input Processing & Indirect Prompt Inj. & Hidden retrieved instructions & High & 15 & \cite{zhan2024injecagent, dong2024jailbreaking, li2025security, li2024targeting, ju2024flooding, wu2024wipi} \\
    Input Handling & Jailbreaking & Safety bypass prompts & High & 26 & \cite{christodorescu2025systems, he2025comprehensive, ghosh2025agentic, feng2025emerged, yu2024llmvirus, choi2025review, edwards2024agent} \\
    Multimodal Input & Adversarial Pert. & Modified images/audio inputs & Med. & 42 & \cite{shayegani2023survey, evertz2024whispers, gan2024navigating, liu2025advances, pan2025measuring, perez2022ignore} \\
    
    \midrule
    \multicolumn{6}{l}{\textit{BRAIN LAYER}} \\
    \midrule
    Planning Module & Goal Hijacking & Manipulating objectives & High & 24 & \cite{datta2025agentic, janjusevic2025hiding, ren2025gtm} \\
    Reasoning Engine & CoT Manipulation & Corrupting reasoning chains & Med. & 21 & \cite{evertz2024whispers, gan2024navigating, mo2024trembling} \\
    Memory System & Memory Poisoning & Injecting false context & High & 41 & \cite{dong2024jailbreaking, ju2024flooding} \\
    Knowledge Base & RAG Poisoning & Corrupting retrieval DB & High & 23 & \cite{li2024targeting, ju2024flooding} \\
    
    \midrule
    \multicolumn{6}{l}{\textit{ACTION LAYER}} \\
    \midrule
    Tool Interface & Tool Manipulation & Unintended tool calls & High & 54 & \cite{barua2024exploring, zhan2024injecagent, debenedetti2024agentdojo, ferrag2025prompt} \\
    API Gateway & Function Hijacking & Redirecting API calls & High & 7 & \cite{janjusevic2025hiding, ferrag2025prompt} \\
    Code Executor & Code Injection & Malicious code execution & Crit. & 13 & \cite{deng2025ai, dornaika2025agentic, triedman2025multi} \\
    Sandbox & Sandbox Escape & Breaking isolation boundaries & Crit. & 8 & \cite{patlan2025real, triedman2025multi} \\
    
    \midrule
    \multicolumn{6}{l}{\textit{INTERACTION LAYER}} \\
    \midrule
    Agent-Agent & Message Injection & Tampering inter-agent comms & High & 12 & \cite{pedro2023prompt, datta2025agentic, deng2025ai, gan2024navigating} \\
    Agent Registry & Impersonation & Masquerading as trusted agent & Med. & 5 & \cite{narajala2025securing, zhu2025master} \\
    \bottomrule
  \end{tabularx}
  
\vspace{6pt}
  \begin{minipage}{\textwidth}
    \footnotesize
    \textit{Note:} The numbers represent the counts at the level of each component. A single paper can contribute to multiple counts because it may discuss more than one type of vulnerability or architectural component. Therefore, the numbers in this table exceed the size of the overall corpus (N = 85) and should not be interpreted as representing individual papers. This column is \textbf{not} the number of unique papers. \\
    
    The severity levels are defined qualitatively using a scheme similar to CVSS v3.1. Critical refers to system compromise or irreversible state changes, High refers to unauthorized data access, goal hijacking, or persistent memory corruption, and Medium refers to behavioral or output manipulation without system compromise. These levels are determined based on worst-case scenarios described in the literature.
  \end{minipage}
\end{table*}

\begin{figure}[!htbp]
  \centering
  \includegraphics[width=0.5\linewidth]{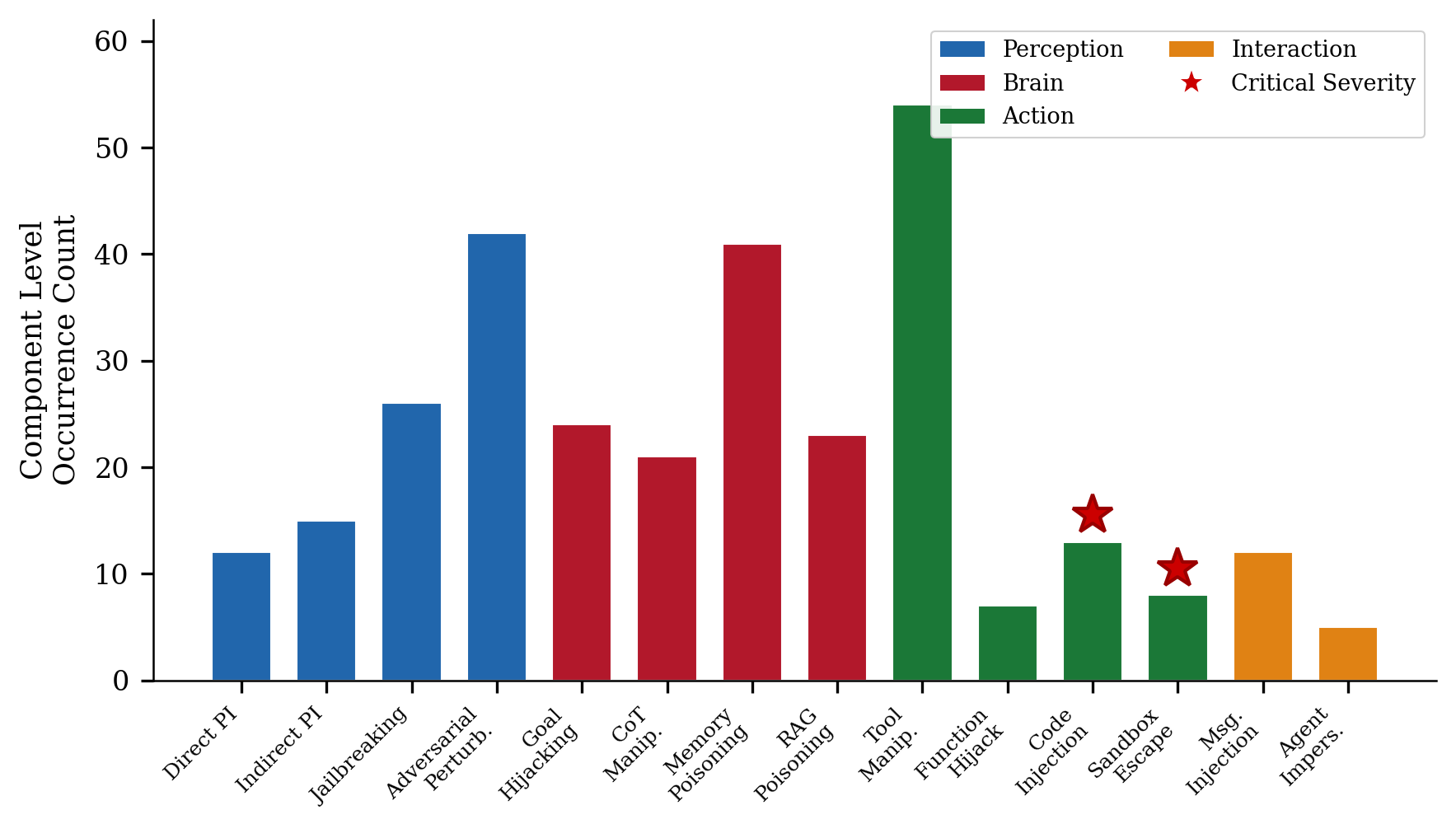}
  \caption{Component-level vulnerability counts and perceived severity distribution.}
  \label{fig:component_severity}
\end{figure}

\subsection{Multi-Dimensional Quantitative Analysis}

This analysis of 85 papers in the vulnerability taxonomy indicates that the following patterns can be identified:

\textbf{Architectural Layer Distribution} Perception layer security risks are the most frequently represented security risks in literature, with 56 out of 85 papers (66\%) focusing on input layer security risks, whereas 35 papers (41\%) discuss brain layer security risks, 4 papers (4.7\%) discuss action layer security risks, and 24 papers (28\%) discuss interaction layer security risks.

This level of concentration appears to be an artifact of methodological accessibility rather than actual risk distribution. Attacks at the perception layer are typically black-box attacks requiring only API access and do not require the development of fully agentic systems. Moreover, early security research on large language models exhibited a path dependency, primarily focusing on prompt manipulation. There was also a bias toward perception-layer attacks due to benchmark availability, as datasets for prompt injection and jailbreaking were developed earlier than those for agentic systems. This does not imply that perception-layer attacks are more dangerous; as shown in Section 5.4, action-layer attacks can be more impactful despite receiving less research attention.

\begin{figure}[!htbp]
  \centering
  \includegraphics[width=0.5\linewidth]{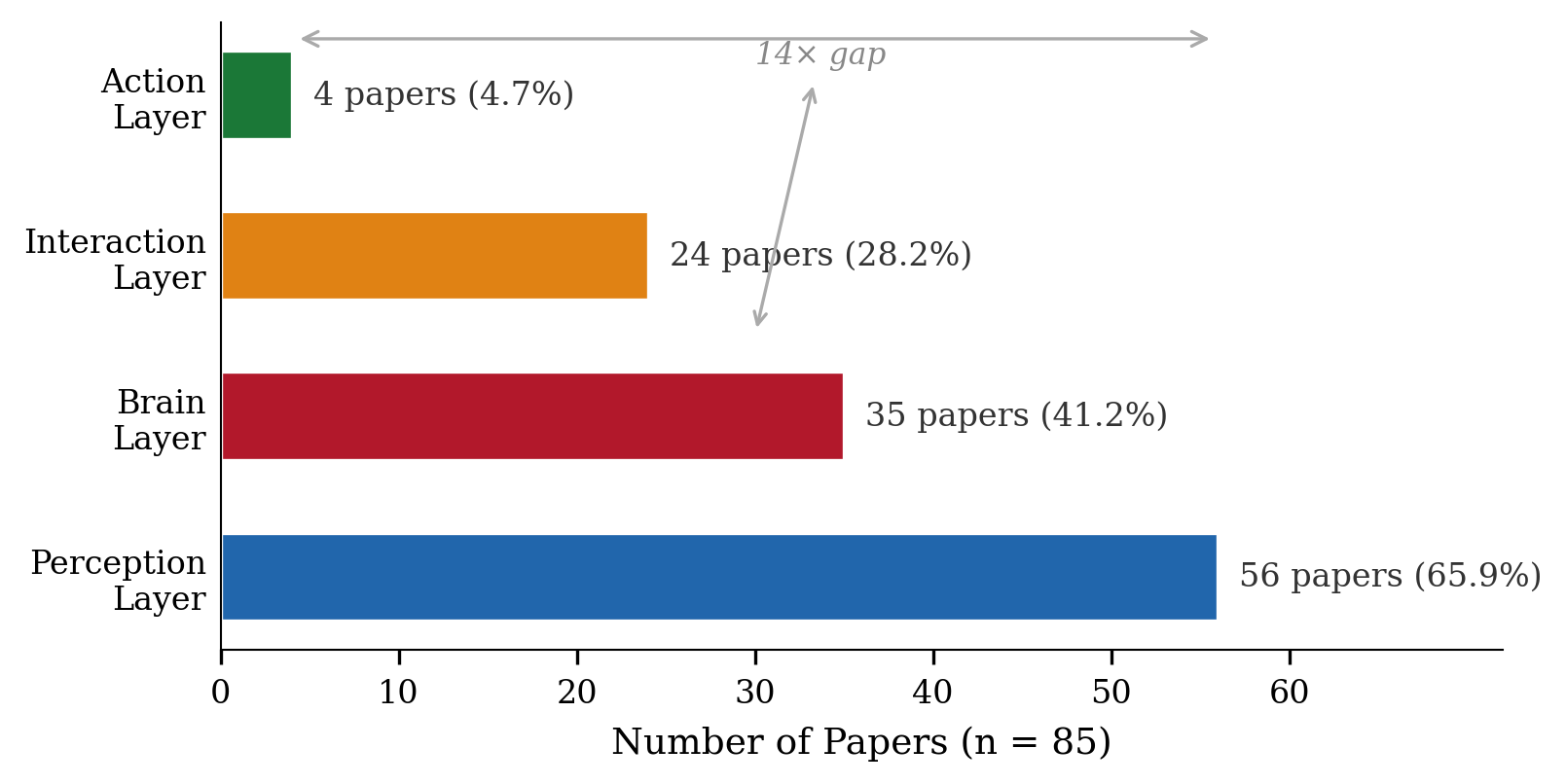}
  \caption{Distribution of vulnerability research across agentic architectural layers.}
  \label{fig:layer_distribution}
\end{figure}

\textbf{Architecture-Specific Patterns} Multi-agent systems being the most frequently represented, with 39 out of 85 papers (46\%) focusing on multi-agent architectures, followed by RAG-based agents with 25 out of 85 papers (29\%), tool-augmented agents with 10 out of 85 papers (12\%), web agents with 4 out of 85 papers (4.7\%). No papers primarily focus on single-agent architectures, though single-agent settings appear as baselines in several studies (Table 8).

\textbf{Vulnerability Type Coverage} Prompt injection examples are the best represented security risk with 36 papers, followed by Adversarial with 30 papers, and jailbreaking with 16 papers. Other security risks are represented by fewer papers: memory poisoning (3), code injection (3), and sandbox escape (2). A large number of papers on these security risks indicate clustering in the literature.

These attacks can be explored in a single-turn setting using standard language model APIs. This contrasts with code injection (3 papers) and sandbox escape (2 papers), which require full agent execution environments with realistic tool integrations. This may indicate that modern research trends are influenced by convenience and not necessarily by risk considerations.

\begin{figure}[!htbp]
  \centering
  \includegraphics[width=0.5\linewidth]{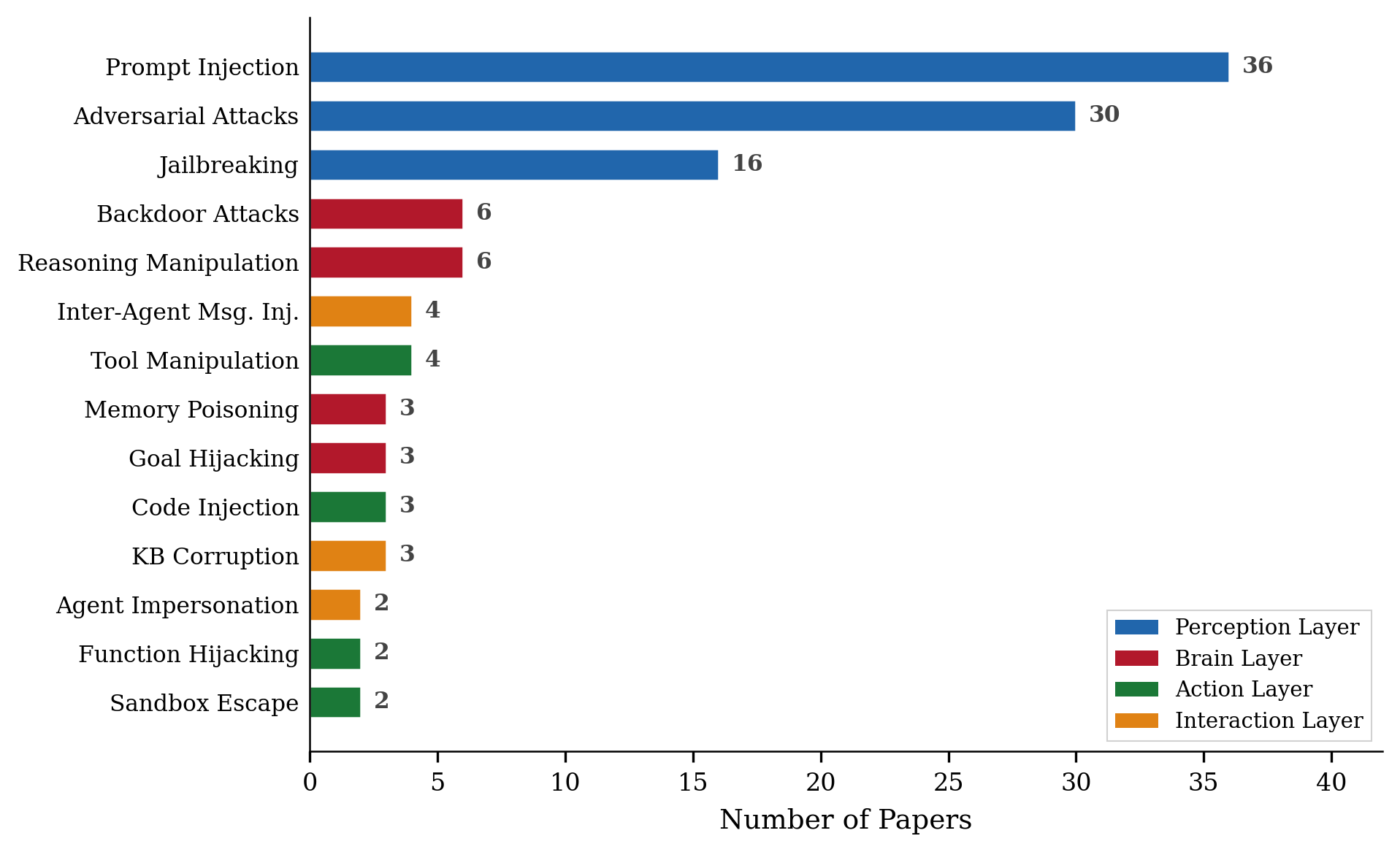}
  \caption{Research frequency by primary vulnerability type.}
  \label{fig:vuln_types}
\end{figure}

\section{Detection Mechanisms in Agentic Systems}

\subsection{Detection Approaches Overview}

Section 4 analyzed the vulnerability landscape, while this section focuses on detection and identifying vulnerabilities. The results show that there is a significant imbalance, with much more research on attack methods than on defense/mitigation-focused work. To be more specific, 47 research papers (55.3\%) focus on vulnerability discovery and attacks, while only 12 research papers (14.1\%) focus on defense and mitigation, a ratio of 3.9:1. 

In terms of agentic large language models, there are a number of factors that contribute to this trend. First, there is a natural incentive structure that favors novel discoveries of attacks rather than incremental improvements to defenses. Second, many attack approaches can be evaluated using only API-level access, while evaluating defenses requires agents that integrate tools and environments. Third, there is currently a lack of benchmarking and metrics for defenses, making it difficult to compare them against one another.

The methods used for detection can be classified under five different stages: input-level detection, runtime detection, multi-agent detection, formal verification, and benchmark-based evaluation. Interestingly, papers addressing action-layer vulnerabilities exhibited lower reproducibility scores (mean 3.1) compared to perception-layer studies (mean 3.9), suggesting that higher-risk architectural layers also suffer from weaker methodological maturity.

\begin{figure}[!htbp]
  \centering
  \includegraphics[width=0.3\linewidth]{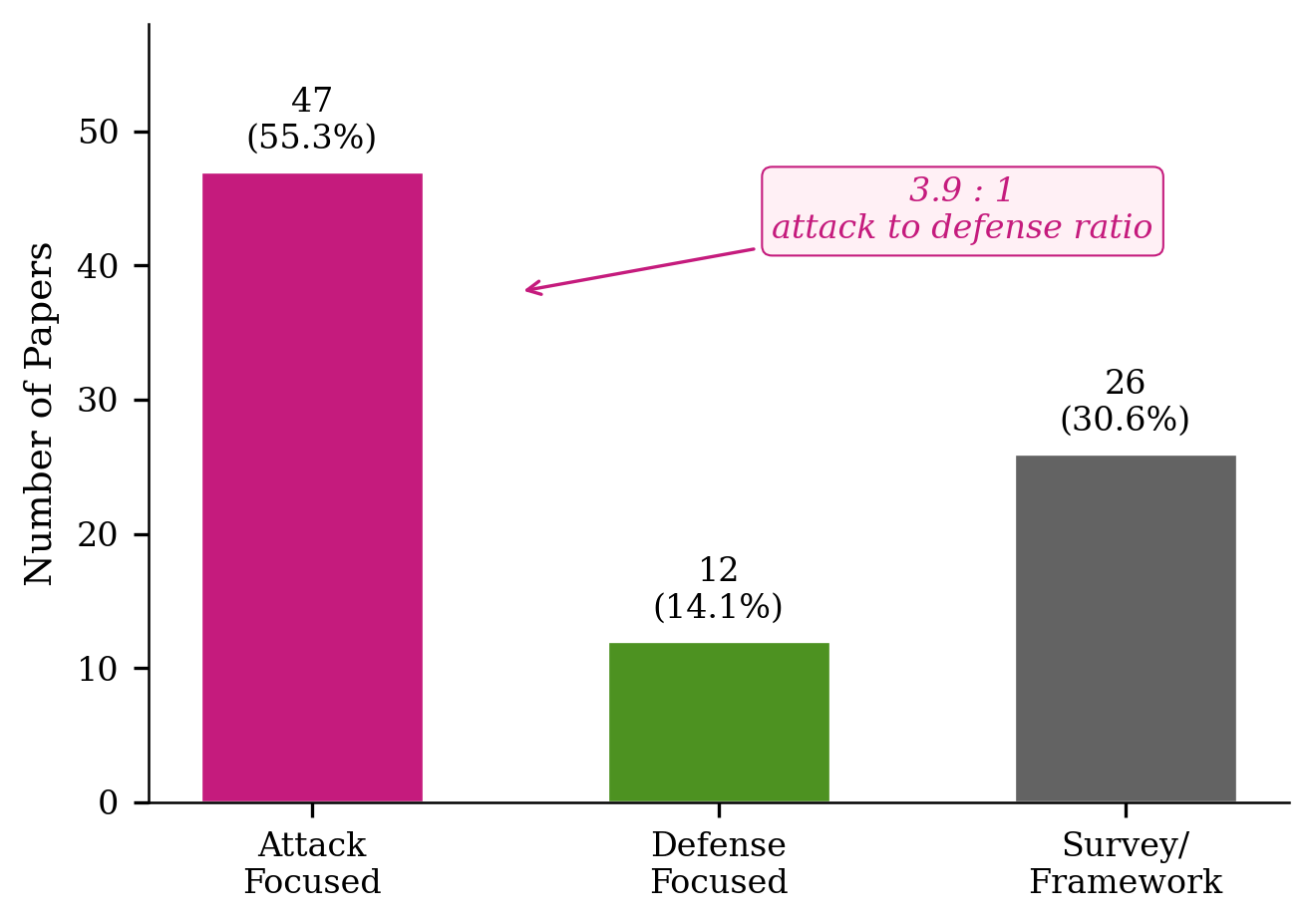}
  \caption{Ratio of attack-focused research to defense/detection-focused publications.}
  \label{fig:attack_defense_ratio}
\end{figure}

\subsection{Input-Level Detection Methods}

Input-level detection is performed before the agent's reasoning module receives input from the user or external environment \cite{zixuan_huang_01f5dea9}. This is, therefore, the first opportunity to identify malicious input. Several approaches have been proposed for prompt-injection detection, including pattern-based and semantic-based detection, as well as classifiers trained on large language models \cite{mengxiao_wang_bf8eaadc}. For instance, known-answer-based detection uses auxiliary models trained on large language models to identify inconsistencies during response generation. This approach relies on the complexity of the attack \cite{erfan_shayegani_a38f4cf2}.

Apart from that, classifier-based detection models are also trained on a set of adversarial input prompts to classify between malicious and benign instructions \cite{shang_wang_b382ed5c}. The classifiers can vary from fine-tuned LLM-based detectors to lightweight embedding-based classifiers that check for deviations from expected input prompt structures \cite{huan_lin_ca227af8}. Embedding-based anomaly detection methods use similarity changes in vector space representations, detecting inputs that significantly differ from expected input distributions \cite{ruiyao_xu_f8b2fa91}. Jailbreak signature detection methods, on the other hand, attempt to detect input patterns associated with instruction overrides, such as role redefinition and policy negation \cite{jiabao_ji_bb95e75e}. The known answer detection method is an extension of the existing detection model, utilizing additional LLM-based instances that generate expected responses and compare them with actual responses, indicating inconsistencies that could imply manipulation \cite{shangbin_feng_c1f34436}. 

In terms of empirical evaluation of input level detection approaches, it is common to measure their performance in terms of reduction in attack success rate, precision/recall of malicious prompts, or robustness of detectors in handling paraphrased attacks \cite{xiaoyu_zhang_1985f4eb}. There have been several works that have evaluated the performance of detectors using jailbreak datasets, adversarial prompts, or injection benchmarks \cite[]{zhen_shang_d199d387, yuqi_jia_dbe9db95, milad_nasr_8e24b340, zhongxing_liao_7c963262}. However, in most common experimental settings, it is often assumed that interactions occur in single turns, i.e., attacks can be contained in a single prompt \cite{hanbin_hong_fbdc6b1f}. This may not be true in multi-turn scenarios, where adversaries can spread their attacks across multiple turns of conversations. In addition, it has also been observed that evaluation of robustness in handling adaptive adversaries, where adversaries can update their prompts based on feedback from detectors, is often not considered \cite{milad_nasr_8e24b340}.

\subsection{Runtime Detection Methods} 

The runtime detection method focuses on detecting agent behavior during runtime, which may indicate an attack \cite{christian_schroeder_de_witt_a4a27db7}.

Behavior monitoring focuses on monitoring agent behaviour to detect and recognize anomalous patterns, including unexpected sequences of tool invocations, resource utilization, or output that does not conform to expected patterns \cite{he_xu_d9a6e601}.

Output verification includes methods that verify agent output using good examples or auxiliary verification methods. In the In-Context Adversarial Game, for example, this method was shown to be highly effective when using the language model as a verifier, as described by \cite{iddo_drori_060eca80}.

In empirical evaluations, mechanisms of runtime detection typically measure the performance of anomaly detection methods in terms of sequences of tool invocation, resource consumption, and consistency of outputs \cite{zhuoyi_yang_d9faa9f3}. The literature has demonstrated better robustness, especially when compared with static prompt filtering, and is more applicable to agents with tool usage. However, there is associated computational cost, especially with heuristics, which can be vulnerable to attacks with multiple steps \cite[]{honglin_mu_663ba1ae, devansh_pandya_3f09bc97, peter_yong_zhong_4eb4649f}. Additionally, there is a high likelihood of evaluations being conducted under controlled conditions, as opposed to an adaptive adversarial framework.

Unlike the approaches of the input level, it is clear from the reviewed corpus that there are no approaches that directly tackle the issue of anomaly detection at the level of execution. The closest approaches are based on the monitoring of tool usage behavior, as presented in one of the reviewed papers. However, it is clear that there is a lack of constraint and boundary enforcement for the level of privilege. Anomalies in memory usage are also a less-explored area. 

The challenges of multi-agent detection include differentiating coordination anomalies from task specialization, monitoring the spread of trust, and checking the integrity of inter-agent communications. Current solutions address these problems in a very limited way. 

\subsection{Evaluation Frameworks and Benchmarking Practices}
No standardized evaluation framework for agentic LLM security currently exists \cite[]{timothy_r__mcintosh_855c6662, hanrong_zhang_3eab6154}. Most studies report attack success rate (ASR), precision/recall of anomaly classification, and in some cases latency overheads \cite[]{yi_dong_7b8a27ed, zhongxing_liao_7c963262, jing_cui_48ebfa5d}. Adaptive adversarial testing, where attackers iteratively refine prompts against detection systems, remains rare \cite[]{yuqi_jia_dbe9db95, milad_nasr_8e24b340}, limiting the comparability of methods across the field \cite{liam_dugan_584410ca}.



\begin{table*}[t]
\centering
\small
\setlength{\tabcolsep}{3pt}
\renewcommand{\arraystretch}{1.1}
\caption{Overview of Agent Security Benchmark Datasets.}
\label{tab:agent_security_datasets}
\resizebox{\textwidth}{!}{
\begin{tabular}{|c|l|c|l|l|l|l|l|}
\hline
\textbf{Ref} & \textbf{Dataset} & \textbf{Year} & \textbf{Size} & \textbf{Vulnerability Types} & \textbf{Agent Scenarios} & \textbf{Attack Types} & \textbf{Availability} \\ \hline
\cite{hanrong_zhang_3eab6154}  & Agent Security Bench (ASB) & 2024 & Multiple scenarios & Prompt Injection, Backdoor & Tool-using, RAG & Direct, Indirect & Public \\ \hline
\cite{zhan2024injecagent}  & InjecAgent & 2024 & 1,054 test cases & Indirect Prompt Injection & Tool-integrated & Indirect & Public \\ \hline
\cite{debenedetti2024agentdojo} & AgentDojo & 2024 & 97 tasks, 629 injections & Prompt Injection & Tool-using & Direct, Indirect & Public \\ \hline
\cite{mazzone2024privacy} & AgentHarm & 2024 & 440 behaviors & Harmful behaviors & General agents & Behavioral & Public \\ \hline
\cite{li2025mind} & AdvWeb & 2024 & Web scenarios & Adversarial attacks & Web agents & Adversarial & Public \\ \hline
\cite{pavlova2024goat} & GOAT & 2024 & Multiple attacks & Jailbreaking & General & Jailbreak & Public \\ \hline
\cite{cui2025mad} & MAD-Spear Dataset & 2025 & Custom & Prompt Injection & Multi-Agent Debate & Social Engineering & Custom \\ \hline
\cite{yang2024security} & Mobile Agent Security Matrix & 2024 & 4 attack paths & Multimodal attacks & Mobile agents & Multi-path & Public \\ \hline
\end{tabular}
}
\end{table*}

\subsection{Detection Methods Summary}
If we consider the ways in which detection works, three main patterns are apparent. The majority of the mechanisms focus mainly on the perception layer, with a strong emphasis on input filtering and anomaly detection at the prompt level itself. In comparison, the runtime monitoring tools are still in their early stages of development and do not yet have empirical validation; only a minority of the literature provides formal guarantees and standardized evaluation methods, reflecting an immaturity level when compared with the attack literature.


\begin{table*}[htbp]
\centering
\begin{threeparttable} 
\caption{Mapping between vulnerability types and detection approaches}
\label{tab:detection_mapping}
\begin{tabular}{|l|l|l|}
\hline
\textbf{Vulnerability Type} & \textbf{Detection Approach} & \textbf{Coverage} \\
\hline
Prompt Injection & Input filtering, prompt validation & Well-studied \\
Adversarial Input & Input sanitization, robustness checks & Well-studied \\
Jailbreaking & Output filtering, policy checks & Well-studied \\
Indirect Prompt Injection & Context validation & Moderate \\
Goal Hijacking & Behavior monitoring & Limited \\
Memory Poisoning & Retrieval validation & Limited \\
RAG Poisoning & Knowledge filtering & Moderate \\
Tool Manipulation & Tool-use monitoring & Limited \\
Code Injection & Execution monitoring & Limited \\
Sandbox Escape & None identified & None \\
Message Injection & Communication validation & Limited \\
Agent Impersonation & Context-based identity checks & Limited \\
Environment Manipulation & Input validation & Limited \\
\hline
\end{tabular}
\begin{tablenotes}[flushleft]
    \small
    \item[] \textit{Coverage indicates the extent of dedicated detection mechanisms identified in the literature: "Well-studied" (multiple established approaches), "Moderate" (some targeted methods), "Limited" (few or indirect approaches), and "None" (no dedicated detection mechanism identified).}
\end{tablenotes}
\end{threeparttable}
\end{table*}

\begin{table*}[t]
\centering
\small
\setlength{\tabcolsep}{3pt}
\renewcommand{\arraystretch}{1.1}
\caption{Overview of Agent Defense Detection Mechanisms}
\label{tab:agent_defense_mechanisms}
\resizebox{\textwidth}{!}{
\begin{tabular}{|c|c|l|l|l|l|l|l|l|l|}
\hline
\textbf{Ref} & \textbf{Year} & \textbf{Method} & \textbf{Vulnerability} & \textbf{Agent Stage} & \textbf{Threat Model} & \textbf{Automation} & \textbf{Dataset} & \textbf{Metrics} & \textbf{Performance} \\ \hline
\cite{ferrag2025prompt} & 2025 & Known-Answer Detection (KAD) & Prompt Injection & Perception & Black-box & LLM-based & Custom & Accuracy, FPR & Variable \\ \hline
\cite{ayzenshteyn2024sentinel} & 2024 & LLM Agent Sentinel & Adversarial & Perception & Black-box & LLM-based & Custom & F1, Accuracy & 95\%+ \\ \hline
\cite{raza2025trism} & 2025 & Reverse Thinking + Role Examination & Jailbreaking & Perception & Black-box & Hybrid & Custom & Detection Rate & Effective \\ \hline
\cite{zhou2024defending} & 2024 & In-Context Adversarial Game (ICAG) & Jailbreaking & Brain & Gray-box & LLM-based & Custom & ASR, Defense Rate & High \\ \hline
\cite{cao2024g4d} & 2024 & Dynamic Guidance Defense (G4D) & Jailbreaking & Perception & Black-box & Hybrid & Benchmark & ASR Reduction & Significant \\ \hline
\cite{zeng2024autodefense} & 2024 & AutoDefense Multi-Agent System & Jailbreaking & Runtime & Black-box & Multi-Agent & JailbreakBench & ASR, FPR & Reduced ASR \\ \hline
\cite{perez2022ignore} & 2023 & Adversarial Prompt Shield & Adversarial & Perception & Black-box & ML-based & Custom & F1, Precision & High \\ \hline
\cite{wu2025slip} & 2025 & Soft Label + Key-Extraction CoT & Prompt Injection & Perception & Black-box & LLM-based & Custom & Detection Rate & Effective \\ \hline
\cite{ferrag2025prompt} & 2025 & Comparative Vulnerability Assessment & Prompt Injection & System & White-box & Manual & Custom & Vuln Count & MCP vs FC \\ \hline
\cite{deng2025ai} & 2025 & Adaptive Encryption + Differential Privacy & Adversarial & System & Gray-box & Hybrid & Framework & Security Metrics & Framework \\ \hline
\end{tabular}
}
\end{table*}

\subsection{Structural Limitations of Current Detection Paradigms}

Despite the progress in detection research, there are several structural gaps in the literature when viewed collectively \cite{david_j__miller_91438ca4}. Firstly, most of these detection methods are situated in the perception layer and are limited in their ability to identify anomalies in prompt injection attacks. Such approaches are generally effective in detecting known types of attacks but are not sustainable in nature \cite{andrew_yeo_2689d92d}.

Secondly, most of these systems rely on additional large language models for evaluating the output of primary agents (LLM-as-judge). Such approaches are promising in nature but are circular in their design and fail to account for robustness against adversarial attacks on the detector \cite{tianchun_wang_2a0bbe08}.

Thirdly, most of these systems are tested in limited conditions and are not comprehensive in their design. Most of these systems are tested in single-session conditions and not in more complex conditions \cite{robin_buchta_50a3edb7}. There is little evidence of robust testing of these systems in long-term conditions or in conditions where tasks are multi-step in nature.

Lastly, most of these systems lack formal guarantees and are limited in their design. Such systems are generally tested empirically and lack formal verification \cite{zhijie_wang_be6dc85c}. Such systems are likely in their design and have not moved towards formalizing unified principles of detection.

Overall, these gaps indicate that the field of detection in large language models has not yet reached methodological maturity and that further research is required in this area \cite[]{simon_valentin_4a84e796, sara_abdali_ce0efbe9, dan_shi_99b59f3c}. Pushing beyond these countermeasures will likely require standardized benchmarks and adaptive threat modeling.

\section{Mitigation Strategies and Defenses}

\noindent To synthesize the architectural distribution of attacks and defenses in agentic LLM systems, we present a unified layered security stack. This model highlights research coverage imbalance and defense maturity across system layers.

\begin{figure}[t]
  \centering
  \includegraphics[width=0.65\linewidth]{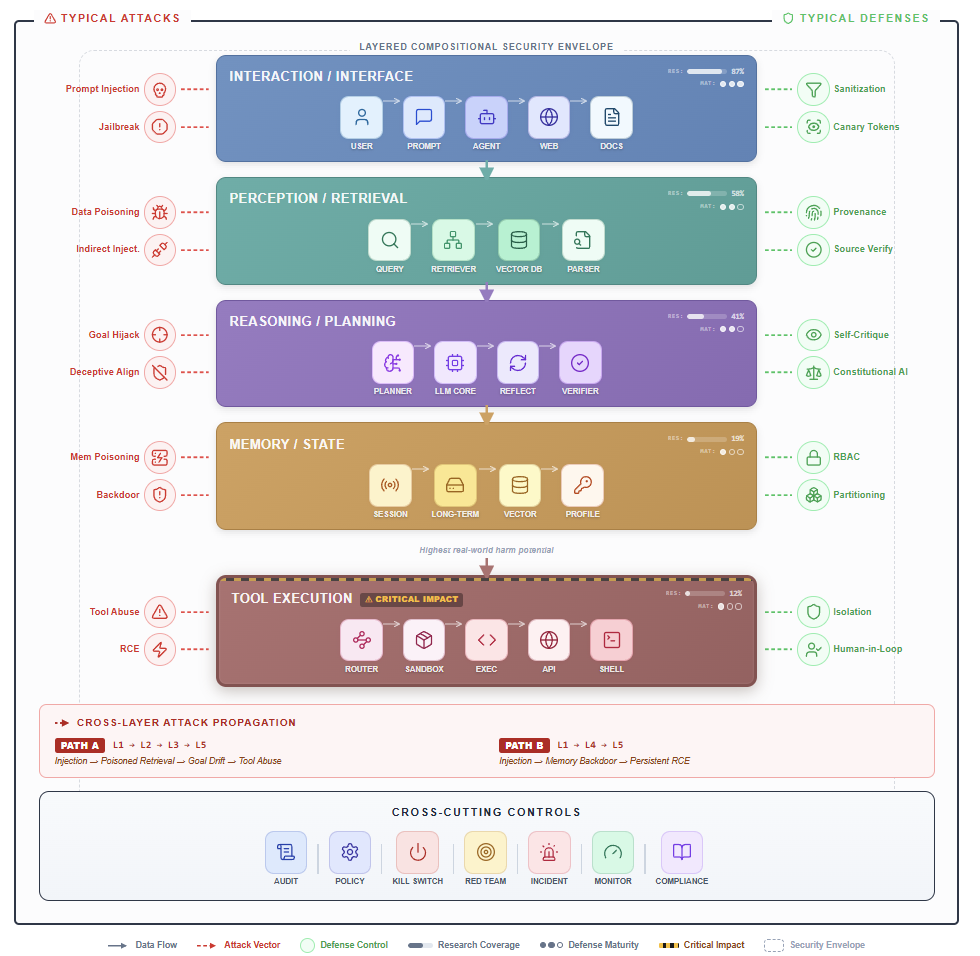}
  \caption{Unified Security Stack for Agentic LLM Systems}
  \label{fig:unified_security_stack}
\end{figure}

\subsection{Defense Taxonomy Overview}

There are six broad categories of defense mechanisms, differentiated by their implementation approach and scope. Detection focuses on identifying vulnerabilities, whereas mitigation aims to reduce their impact \cite[]{mohamad_fazelnia_b030670d, benji_peng_910433b5}. Our review identified 17 defense-focused studies (12 detection and 5 mitigation-only), with emphasis on filtering (60\%), runtime control (70\%), and multi-agent strategies (35\%); note that these percentages are not mutually exclusive, as individual studies may employ multiple defensive mechanisms. Categories were assigned based on each paper's primary defensive mechanism and consolidated into stable groups: input filtering, architectural modification, runtime control, training-time mitigation, governance, and multi-agent defense.

\subsection{Input/Output Filtering and Shields}
Input/Output filtering primarily addresses perception-layer vulnerabilities (Section 5.2), particularly direct and indirect prompt injection. Input filtering is the process of removing or modifying bad inputs before they enter the process stage \cite{qendro__lorena_dbae4277}. Prompt sanitization is also mentioned in the literature \cite{simon_ostermann_0adcf5ae}. There are studies that introduced input prompts that filter out attack patterns that come in the form of bad input. Guard models have been proposed for input filtering, screening out malicious inputs before they reach the reasoning module. For example, Auto Defense uses a multi-agent filtering system to filter inputs through specialized screening agents that identify prompt injection and unsafe instructions before passing them to the main agent as outputs.

Output filtering is also addressed; the outputs are filtered before they are released to other systems \cite[]{valentyn_boreiko_a6cfde2c, yuntao_wang_8d255dc9}. This is especially important for code-execution agents; filtered outputs prevent malicious code execution.

Generally, empirical evaluations of the efficacy of the filtering mechanisms for input and output filtering demonstrate a reduction in the attack success rate (ASR) for curated datasets for jailbreak and prompt injection attacks \cite{y__m__wang_1f741a81}. However, the efficacy of the guard model filtering approach, while significant, demonstrates a reduction in efficacy against adaptive threat actors, particularly those employing alternative definitions for malicious intent, such as distributing instructions across multiple turns in the conversation \cite{p__kulkarni_4930b9be}. Output filtering for code execution agents significantly mitigates the risk of direct execution but does not account for semantic harm embedded in code, particularly if the code is syntactically valid \cite{andrei_kucharavy_3559fecb}. However, overly aggressive filtering can result in false positives, reducing the utility of the task. 

\subsection{Architectural Defenses}
Architectural defense mechanisms primarily address brain-layer vulnerabilities (Section 5.3), particularly through the addition of control points between reasoning (B) and execution (E). The architectural techniques focus on modifying the agent system to prevent potential attacks \cite{tri_cao_cef04a89}. 

\textbf{Layered Agent Architectures} Layered architectures are proposed for agents; they separate reasoning and execution via intermediate verification steps \cite{christian_schroeder_de_witt_a4a27db7}.

AutoDefense proposes multi-agent systems where one agent monitors another, detecting and filtering suspicious behaviour \cite{yuntao_wang_bc05705f}.

\textbf{Privilege Separation} Limiting agent capabilities through permission systems, capability-based security, and role-based access control \cite{kaiyuan_zhang_1c830989}.

Architectures that support multi-agent monitoring and cross-agent 
verification tend to exhibit improved robustness against simple 
jailbreak scenarios; however, they impose additional communication 
overhead and risk correlated failure when identical models are used 
across agents \cite{aryan_bhatt_37977ac2}. Privilege separation 
limits worst-case damage potential but does not prevent malicious 
strategy development within defined limits. As architectural 
complexity increases, unintended attack surfaces may also grow, 
particularly in multi-agent scenarios where inter-agent communication 
becomes an attack vector \cite{sudhir_shrestha_8e892138}.

\subsection{Runtime Controls}
Runtime control mechanisms primarily address action-layer vulnerabilities (Section 5.4), particularly through execution constraint.

Runtime controls monitor and constrain agent execution \cite{herman_errico_b4144761}.

\textbf{Sandboxing and Isolation} Executing agent code in isolated environments with limited system access.

Sandboxing is particularly important for code-execution agents but applies to all tool-using agents.

\textbf{Rate Limiting and Quotas:} Constraining resource consumption and tool invocation frequency to limit the impact

of compromised agents.

\textbf{Policy Engines} Enforcing declarative policies about which tools agents can invoke and under what conditions and circumstances.

The control mechanisms can be evaluated with respect to their effectiveness in containing scenarios such as simulated privilege escalation, unauthorized API calls, and resource exhaustion attacks \cite{david_ferraiolo_67b0032e}. The mechanisms of sandboxes/isolation show high levels of containment, especially with respect to code execution agents, as they restrict the impact of malicious outputs even if upstream reasoning is compromised. Rate limiting reduces the impact of compromised agents but cannot prevent a high-impact action \cite{xuhui_zhou_3ea3e7cb}. The policy engines show high levels of declarative enforcement but are heavily dependent on the precise definition of interactions between tools. 

Empirical studies confirm that sandboxing is effective at the execution level, though per-action policy validation introduces latency overhead and misconfiguration risk.

\subsection{Training-Time Mitigation}

Training-time mitigation primarily addresses brain-layer vulnerabilities (Section 5.3), particularly through control of statistical behavior of reasoning operator (B).
Training-time strategies address the issue during model training. Adversarial training makes the model more robust by training it on adversarial examples \cite{nandish_chattopadhyay_433f3644}. This allows the model to learn how to be robust against existing attack patterns. Reinforcement Learning from Human Feedback (RLHF) promotes safe behavior by incorporating human feedback into the reinforcement learning optimization loop. This discourages unsafe behavior in the model---techniques for backdoor remediation focus on detecting backdoors via fine-tuning or cleansing.

Adversarial fine-tuning helps improve robustness against known attacks, and reinforcement learning from human feedback helps prevent harmful outputs, but these methods do not work for new patterns of attacks. Backdoor remediation helps prevent risks of unknown triggers but involves some level of performance degradation.

\subsection{Governance and Oversight}
Governance mechanisms primarily address interaction-layer vulnerabilities (Section 5.5), particularly through control of interactions between agents and external systems. 

In addition to technical measures, process-level defenses can be implemented by involving humans in the operations of agents \cite{bang_liu_60553ab1}. Such measures, which involve human-in-the-loop, require authorization from humans before executing certain operations, thereby providing a highly reliable security mechanism at the expense of reduced scalability \cite{tobin_south_fa1dd572}. Audit logging can improve accountability by recording all operations performed by agents, enabling investigations of security breaches.

The reliability of human-in-the-loop mechanisms in preventing catastrophic behaviors is very high, especially in high-risk domains \cite{shuiguang_deng_3e6960c4}. However, empirical studies on the deployment of these mechanisms indicate that there are issues with scalability as well as increased latency in multi-step agent-based workflows. The reliance on human oversight is also conditional on the ability of the human operator to identify the subtle changes in the reasoning process, which is not always possible \cite{tula_masterman_0f46e5c1}. 

\subsection{Multi-Agent Specific Defenses}
Multi-agent-specific defenses primarily address interaction-layer vulnerabilities (Section 5.5), particularly those arising from inter-agent communication and coordination. There are several defensive concerns that apply to multi-agent systems. In relation to communication security, cryptographic signing is used to prevent tampering and spoofing \cite{christian_schroeder_de_witt_a4a27db7}. In relation to consensus, agreement among several agents is required before specific actions are taken. In relation to agent authentication, verification of agent identities is required to prevent impersonation attacks. Table~\ref{tab:vuln-detect-mitigate} synthesizes the relationship between vulnerability types (Section~5), detection mechanisms (Section~6), and mitigation strategies (Section~7), highlighting remaining coverage gaps.

\begin{table*}[t]
\centering
\caption{Vulnerability--Detection--Mitigation Mapping across Agentic Systems}
\label{tab:vuln-detect-mitigate}
\small
\begin{tabular}{p{3.2cm} p{3.5cm} p{3.5cm} p{4cm}}
\toprule
\textbf{Vulnerability Type} & \textbf{Detection Approach(es)} & \textbf{Mitigation Approach(es)} & \textbf{Coverage Gap} \\
\midrule

Prompt Injection (Perception) 
& Guard models, input anomaly detection, LLM-as-judge 
& Input/output filtering, prompt sanitization 
& Weak against multi-turn and obfuscated attacks \\

Data Poisoning (Perception) 
& Data validation, provenance tracking 
& Dataset filtering, training-time mitigation 
& Limited real-time detection capabilities \\

Reasoning Manipulation (Brain) 
& Chain-of-thought inconsistency detection 
& Architectural defenses, training-time alignment 
& Difficult to detect semantic manipulation reliably \\

Tool Misuse (Action) 
& Execution monitoring, policy violation detection 
& Runtime controls, sandboxing, policy engines 
& Cannot fully prevent high-impact actions \\

Multi-Agent Message Injection (Interaction) 
& Communication auditing, anomaly detection 
& Multi-agent verification, governance mechanisms 
& Vulnerable to coordinated adversarial agents \\

\bottomrule
\end{tabular}
\end{table*}

\begin{table}[t]
\centering
\caption{Trade-offs across mitigation strategies in agentic systems}
\label{tab:mitigation-tradeoffs}
\small
\begin{tabular}{p{2.8cm} p{2cm} p{2cm} p{2cm} p{2.3cm}}
\toprule
\textbf{Strategy} & \textbf{Effectiveness} & \textbf{Scalability} & \textbf{Overhead} & \textbf{Complexity} \\
\midrule

Input/Output Filtering 
& Moderate (weak vs adaptive attacks) 
& High 
& Low 
& Low \\

Architectural Defenses 
& High (structural guarantees) 
& Moderate 
& Moderate 
& High \\

Runtime Controls 
& High (strong containment) 
& Moderate 
& High (latency, monitoring) 
& Moderate \\

Training-Time Mitigation 
& Moderate (known attacks) 
& High 
& Low (runtime) 
& Moderate \\

Governance (Human-in-the-loop) 
& Very High (high assurance) 
& Low 
& High (human cost) 
& Moderate \\

Multi-Agent Defense 
& Moderate--High 
& Moderate 
& Moderate--High 
& High \\

\bottomrule
\end{tabular}
\end{table}



\begin{table*}[t]
\centering
\small
\setlength{\tabcolsep}{3pt}
\renewcommand{\arraystretch}{1.1}
\caption{Overview of Agent Defense Strategies and Implementation}
\label{tab:agent_defense_strategies}
\resizebox{\textwidth}{!}{
\begin{tabular}{|c|l|l|l|l|l|l|l|}
\hline
\textbf{Ref} & \textbf{Category} & \textbf{Technique} & \textbf{Agent Type} & \textbf{Level} & \textbf{Evaluation} & \textbf{Effectiveness} & \textbf{Limitations} \\ \hline
\cite{zeng2024autodefense} & Multi-Agent Defense & LLM-based response filter agents & Multi-Agent & System & JailbreakBench & High & Latency overhead \\ \hline
\cite{zhou2024defending} & Runtime Defense & In-Context Adversarial Game & Single Agent & Model & Custom & High & Training required \\ \hline
\cite{cao2024g4d} & Input Filtering & Dynamic guidance injection & Single Agent & API & Multiple & Significant & May affect utility \\ \hline
\cite{wu2025slip} & Input Filtering & Soft label + CoT defense & RAG-based & System & Custom & Effective & Domain-specific \\ \hline
\cite{deng2025ai} & Architectural & Adaptive encryption + differential privacy & Multi-Agent & System & Framework & Framework & Complexity \\ \hline
\cite{raza2025trism} & Runtime Defense & Agent monitoring + admin intervention & Multi-Agent & System & Custom & Effective & Human overhead \\ \hline
\cite{edwards2024agent} & Training-time & Backdoor trigger removal fine-tuning & Single Agent & Model & Benchmark & Moderate & May affect performance \\ \hline
\cite{fasha2024owasp} & Architectural & OWASP-aligned agent defense framework & Tool-using & System & OWASP & Framework & Implementation effort \\ \hline
\cite{deng2025ai} & Runtime Defense & Runtime security controls for LPCI & Single Agent & System & Custom & Proposed & Not validated \\ \hline
\cite{perez2022ignore} & Input Filtering & Adversarial prompt detection shield & Single Agent & API & Custom & High & False positives \\ \hline
\end{tabular}
}

\vspace{2pt}
\begin{minipage}{\textwidth}
\small \textit{Note:} While Table 5 presents the categorical taxonomy of vulnerabilities, Table 6 quantifies their distribution across the 85 analyzed studies.
\end{minipage}
\end{table*}

\section{Research Challenges and Future Directions}
Based on the quantitative results presented in Sections 5-7 of this chapter, as well as the cross-layer insecurity framework described above, this section attempts to synthesize the identified research gaps in order to outline the containment-oriented research avenues for secure agentic architectures.

\begin{table*}[t]
\centering
\small
\setlength{\tabcolsep}{4pt}
\renewcommand{\arraystretch}{1.1}
\caption{Quantitative Overview of Agentic LLM Security Literature (2023--2025, $n=85$)}
\label{tab:literature_overview}
{
\begin{tabular}{|l|l|l|}
\hline
\textbf{Metric} & \textbf{Value} & \textbf{Notes} \\ \hline

Total Papers Analyzed & 85 & 2023--2025 publication window \\ \hline

Attack-focused Papers & 47 (55.3\%) & Primary focus on vulnerability discovery \\ \hline

Defense-focused Papers & 12 (14.1\%) & 3.9:1 attack-to-defense ratio \\ \hline

Survey/Framework Papers & 26 (30.6\%) & Includes survey/taxonomy/framework studies \\ \hline

Papers from 2025 & 62 (72.9\%) & Rapid field expansion year \\ \hline

Papers from 2024 & 21 (24.7\%) & Field consolidation year \\ \hline

Papers from 2023 & 2 (2.4\%) & Early foundational work \\ \hline

Multi-Agent Focus & 39 (45.9\%) & Most represented architecture \\ \hline

Tool-Augmented Focus & 10 (11.8\%) & Underexplored despite real-world adoption \\ \hline

Code-Execution Focus & 3 (3.5\%) & Critical architectural research gap \\ \hline

Perception-Layer Research & 56 (65.9\%) & Dominant research layer \\ \hline

Action-Layer Research & 4 (4.7\%) & 14x gap vs perception layer \\ \hline

Prompt Injection Papers & 36 (42.4\%) & Most studied vulnerability type \\ \hline

Adversarial Attack Papers & 30 (35.3\%) & Second most studied \\ \hline

Jailbreaking Papers & 16 (18.8\%) & Third most studied \\ \hline

\end{tabular}
}
\end{table*}

\subsection{Open Problems and Critical Gaps}

In accordance with the quantitative analysis provided in Sections 5, 6, and 7, seven critical open problems have been identified. These problems have been listed as follows:

\textbf{Problem 1: Security of Code-Execution Agents (3 papers)}

From the literature, it is evident that there is a significant gap in research on the security of code execution agents. This represents the most critical underexplored area identified in this review. Code injection is a critical security issue that can potentially allow for arbitrary code execution within the context of the code execution agent. The main technical challenge is to design sandbox mechanisms that are sufficiently restrictive to prevent agents from escaping while being sufficiently permissive to allow agents to function as intended. One example is a data analysis agent that generates Python code for handling uploaded data. If the generated code is vulnerable to a path traversal attack or runs shell commands, an adversary with control over the data can steal server files or make unauthorized network connections.

\textbf{Problem 2: Security of Embodied Agents (0 papers)}

Despite the growing deployment of robotic systems, the security of embodied agents remains unaddressed in the surveyed literature. Embodied agents have several security issues, including sensor manipulation and actuator control, with direct implications within the physical domain. The main technical challenge is that agents have a tight coupling between perception and action, which complicates the enforcement of reliable safety constraints due to sensor noise and environmental uncertainties. One example is a robotic warehouse agent where adversary labels placed on objects affect perception and lead to incorrect routing decisions.

\textbf{Problem 3: Security of Single-Agent Systems (0 papers primarily focused)}

Single-agent systems were defined strictly as architectures without tool use, retrieval augmentation, multi-agent interaction, or persistent memory. Under this strict definition, no studies explicitly focused on security analysis of such minimal agent architectures, though single-agent settings appear as baselines in several studies (Table~\ref{tab:literature_overview}). The main technical challenge is that agents have a complex internal reasoning process that complicates isolating and analyzing failure modes without resorting to oversimplification of model behavior. This cluster represents limitations in detection capabilities and evaluation methodologies, especially in adapting to adversarial situations and developing standardized benchmarks.

\textbf{Problem 4: Maturity of Detection Methods (10 detection methods vs 52 attacks)}

There is a 5:1 ratio of attacks to detection methods, thus reflecting a significant imbalance. It is evident that the maturity level of detection methods is relatively low, with limited evaluation of effectiveness and a lack of standardized protocols. The main technical challenge is that adaptive attackers will continually try to evade detection systems and optimize attacks, which makes classifiers and rule-based detection systems only temporarily effective.

\textbf{Problem 5: Coverage of Tool-Augmented Agents (12\%)}

Only 12\% of research literature deals with tool-augmented agents, despite (a) the fast pace at which tool-calling large language models are being deployed in production environments, (b) the special risks tool access imposes, and (c) the large attack surface in enterprise environments. The main technical challenge is to design security mechanisms for tool invocation pipelines that have heterogeneous trust boundaries and side effects.

\textbf{Problem 6: Real-World Deployment}

Most research is conducted in a lab environment, which does not accurately reflect real-world conditions where attackers operate with limited information, communicate asynchronously, and deal with uncontrolled external data.

\textbf{Problem 7: Standardized Evaluation (absent)}

Despite eight benchmark datasets being available, there are no standardized evaluation protocols. The main technical challenge is that there is a lack of a unified evaluation protocol that makes it difficult to compare results and security baselines across different studies.

\subsection{Future Research Agenda}

We outline the following research agenda:

\textbf{Short term (1-2 years):} The focus is on securing code execution and tool-assisted agent augmentation. The emphasis is on filling critical research gaps in high-risk deployment scenarios \cite{nikhil_patnaik_1b63eb23}. The research community should develop an evaluation framework and conduct empirical studies in real-world deployment scenarios, moving beyond laboratory-based research. 

\textbf{Medium term (3-5 years):} The scope of the research should be broadened to include agent embodiment, agent vulnerabilities, formalized detection methodologies, and tool supply chain security \cite{zehang_deng_1ca5d202}. 

\textbf{Long-term (5+ years):} The research community's goal is to achieve formal verification of security properties, trustworthy agent architectures, and human oversight.

\subsection{Standardization Needs}
Advancing agentic LLM security requires standardization across several dimensions: a unified vulnerability taxonomy aligned with the four-layer framework proposed in this survey, standardized evaluation metrics (attack success rate, detection accuracy, false positive/negative rates), a unified benchmark protocol for comparing detection and defense effectiveness \cite{junhao_zheng_8c8fccef}, and guidelines for responsible disclosure of newly identified agent security flaws \cite{benji_peng_910433b5}.

\section{Conclusion}

This systematic literature review examines 85 research articles published between 2023 and 2025, which discuss security issues related to agentic large language model (LLM) systems. The review follows the PRISMA guidelines, applying a rigorous methodology to systematically search databases, apply inclusion/exclusion criteria, and assess the quality of all included research articles. 
We propose a four-layer vulnerability taxonomy that maps thirteen vulnerability types to the perception, reasoning (brain), action, and interaction layers of agentic architectures, providing a classification grounded in architectural function rather than inconsistent attack nomenclature. Our quantitative analysis of eighty-five papers reveals a 3.9:1 attack-to-defense ratio and a fourteen-fold disparity between perception-layer and action-layer research coverage, providing the first systematic assessment of structural imbalance in the field.
Agentic LLM systems represent a significant leap in capability while introducing new security challenges distinct from those of traditional LLMs and AI agents. The findings of this systematic review reveal that, although research on vulnerabilities is growing, work on defense mechanisms remains limited, representing a critical gap. As autonomous agentic LLM applications continue to evolve, the security issues identified in this review are likely to become increasingly important for both researchers and practitioners.

\printbibliography
\end{document}